\documentclass[oldversion]{aa} 
\usepackage{graphicx}
\usepackage{aalongtable}
\usepackage{txfonts}
\usepackage{rotating}
\usepackage{lscape}
\usepackage{placeins}
\newcommand{\gsim}{\;\lower.6ex\hbox{$\sim$}\kern-7.75pt\raise.65ex\hbox{$>$}\;}
\newcommand{\lsim}{\;\lower.6ex\hbox{$\sim$}\kern-7.75pt\raise.65ex\hbox{$<$}\;}

\begin{document}
\title{Discrepancies between spectroscopy and HST photometry in tagging multiple
stellar populations in 22 globular clusters
\thanks{Full Table 3 and Table A.1 in Appendix A are only available at the CDS
via anonymous ftp to cdsarc.u-strasbg.fr (130.79.128.5) or via http://cdsarc.
u-strasbg.fr/viz-bin/cat/J/A+A/??/??}
 }

\author{
Eugenio Carretta\inst{1}
\and
Angela Bragaglia\inst{1}
}

\authorrunning{Carretta and Bragaglia}
\titlerunning{Discrepancies in population tagging}

\offprints{E. Carretta, eugenio.carretta@inaf.it}

\institute{
INAF-Osservatorio di Astrofisica e Scienza dello Spazio di Bologna, Via P. Gobetti
 93/3, I-40129 Bologna, Italy}

\date{}

\abstract{Multiple populations (MPs) in globular clusters (GCs) are stars
distinct by their abundances of light elements. The MPs can be directly
separated by measuring abundances of C, N, O, Na, Al, Mg with spectroscopy or
indirectly from photometric sequences created by the impact of different
chemistry on band passes of particular filters, such as the HST pseudo-colours in
the ultraviolet. An attempt to link HST pseudo-colours maps (PCMs) and spectroscopy was
done by Marino et al. (2019), using abundances mostly from our FLAMES survey.
However, we uncovered that an incomplete census of stars in common was used in
their population tagging. We correct the situation by building our own PCMs and
matching them with our abundances in 20 GCs, plus two GCs from other sources,
doubling the sample with spectroscopic abundances available. We found that the
pseudo-colour $(mag_{F275W}-2 \times mag_{F336W}+mag_{F438W})$ does not have a monotonic trend with Na
abundances, enhanced by proton-capture reactions in MPs. Moreover, on average
about 16\% of stars with spectroscopic Na abundances show a discrepant tagging
of MPs with respect to the HST photometry. Stars with chemistry of second
generation (SG) are mistaken for first generation (FG) objects according to HST
photometry and vice versa. In general, photometric indices tend to overestimate
the fraction of FG stars, in particular in low mass GCs. We offer a simple
explanation for these findings. Finally, we publish all our PCMs, with more than
31,800 stars in 22 GCs, with star ID and coordinates, for easy check and
reproduction, as it should be in science papers.
}

\maketitle

\section{Introduction}

Multiple stellar populations (MPs) in Galactic globular clusters (GCs) are 
commonly defined as ensembles of cluster stars that are distinct by their
content in light elements. Striking differences in C and N, as well as in Na and
O content, among GC stars were known and studied since the '80s (see Smith 1987 and
Kraft 1994 for reviews on those studies).
Since the seminal paper by Gratton et al. (2001) on
the abundances of unevolved stars in two clusters, it is crystal clear that
presently evolving low mass stars cannot be the factories of the observed
chemical pattern. For production of species like N, Na, Al (and sometimes Si, Ca,
Sc, K) combined to simultaneous depletion of C, O, Mg, Gratton et al. indicated
as responsible the most massive stars, now extinct, of a stellar population
that preceded the one presently observed in GCs. The polluters among first
generation (FG) stars contributed matter nuclearly processed by proton-capture
reactions in H-burning at high temperature. Different cycles are involved,
depending on the temperature thresholds achievable according to the polluter
mass. The processed matter, coupled with diluting pristine gas, constitutes the
necessary mass budget required to generate the subsequent second generation(s)
(SG) with
the observed chemical pattern (see e.g. the extensive reviews by Gratton et al.
2004, 2019 and references therein). In contrast to this widely accepted
scenario, there is no consensus on a variety of issues regarding MPs, starting 
from no less than the nature of the polluters required to provide the raw
material to be recycled into subsequent stellar generations or populations with
modified composition. This mainly stems from the difficulties found to match
(quantitatively, and sometimes qualitatively) theoretical models to the observed
pattern of correlations and anti-correlations among light element abundances in
GC stars (see e.g. Bastian and Lardo 2018). Thus, none of the many
candidates proposed for FG polluters (e.g. intermediate mass asymptotic giant
branch, AGB, stars: Cottrell and Da Costa 1981, Ventura et al. 2001; massive,
fast rotating stars: Decressin et al. 2007; massive binaries: de Mink et al.
2009; super-massive stars: Denissenkov and Hartwick 2014) seems to provide a
fully satisfactory answer so far.

To gain more insight on MPs, it is useful to to study and compare the properties
of distinct populations. The bedrock to this approach is to identify them with
precision. In turn, the ratio first to second generations is a key parameter to
effectively constrain models to reconstruct the formation and early evolution of
GCs. As an example, the extensive database of [O/Na] values spanned by different
stellar populations in GCs allowed to define the main property of MPs in GCs,
i.e. that the relevance of the MP phenomenon is strictly correlated to the mass
of the GC (Carretta 2006, figures 12 and 13; Carretta et al. 2010a), a fact
later confirmed also by photometric means (e.g. Milone et al. 2017).

At the same time, the homogeneous tagging of stellar 
populations with primordial and altered chemical composition in a large set of GCs allowed to draw
attention to the so called mass budget problem. Since the chemical composition
of the polluted component must be inherited from only a part (the most massive) of
an early stellar generation, the population ratio has a significant impact on
every scenario devised to explain the formation of GCs, since it provide strong
constraints to the initial GC mass. The proportion of FG and SG (about 30:70) 
stars derived from homogeneous spectroscopy and the Na-O anti-correlation implies
that initially GCs could have been much more massive than today and that most of
their FG low mass stars were lost at early times in the GC lifetime (e.g.
Prantzos and Charbonnel 2006, Decressin et al. 2007, Carretta et al. 2010a,
D'Ercole et al. 2010, Vesperini et al. 2010, 2013).

It is then of paramount relevance to correctly identify among GC stars those
maintaining the original composition, identical to that of field stars of
similar metallicity for all practical purposes, and those belonging to the
next generation born in the cluster.

The main method to tackle this issue is to directly measure the abundances of
light elements with spectroscopy. Spectra, preferentially at high resolution for
accurate abundances of atomic species, but even at low resolution for molecular
features, allow to directly separate stars with FG and SG chemical composition.
Since different species are destroyed or produced at different temperatures, a
complete survey of light element abundances is actually a tomography performed
on the temperature distribution, hence on the  involved mass range, of the
putative polluters.

Indirect measures are possible with photometry, exploiting filters with
bandpasses that cover wavelength ranges affected by change in flux due to the
effect of molecular bands, whose strength is modified by reactions enhancing N at
the espense of depleting C and O. Hence, usually only UV/blue filters,
encompassing the regions where OH, NH, and CN molecular features are located, 
are particularly efficient in this specific task. Several photometric systems 
have been used to this purpose (see e.g. Calamida et al. 2007, Sbordone et al. 2011, Carretta et al.
2011a, Lee et al. 2009, Massari et al. 2016 for $ubvy$
Str\"omgren filters; Monelli et al. 2013 for the Johnson UBVI bands; 
Johnson et al. 2023 for the Sloan $ugri$ system;
Piotto et al. 2015 and Milone et al. 2017 for the combinations using
ultraviolet HST filters). 

Suitable combinations of magnitudes in these filters result in pseudo-colours
able to separate photometric sequences mostly populated by stars of different
composition. The method uses appropriate normalizations suitable to clearly
enhance the separation of populations with different abundance of light
elements, in particular C, N, and sometimes O. 

Some photometric systems seem more suitable than others to provide accurate
population tagging of stars in GCs (see e.g. Lee 2019, 2023), but recently the
approach that uses UV/blue/optical HST photometry has drawn much attention, due
to the large sample of GCs observed (57) and to the exploitation of more
absorbtion molecular bands in some suitable filters: OH in F275W, NH in F336W,
CN and CH in F438W, again NH in F343N (the last was employed by Larsen et al.
2014 to study GCs in the Fornax dwarf spheroidal galaxy).

The largest difference is obtained using a pseudo-colour derived by subtracting
from the F275W magnitude (sensitive to the abundance of O, which is depleted in
proton-capture reaction) two times the F336W magnitude (sensitive to N, which is
enhanced) and finally adding the magnitude in 
F438W (whose bandpass includes region of CH, another species depleted).  Other,
redder filters such as F814W are not particularly sensitive to the MP atmospheric
composition, not including useful regions of molecular bands, but in combination
with filters at much bluer wavelengths (e.g. F275W) are very sensitive to
variations in the star effective temperature. In the assumption that the changes
in light elements are accompanied by small differences in the He  content,
colours like $mag_{F275W}-mag_{F814W}$, combining UV and optical bands, may provide some
indications about  He. 

A few words on notation. For clarity and conciseness we will adopt the
simplifications illustrated in Table~\ref{t:notation}.

\begin{figure*}
\centering
\includegraphics[scale=0.8]{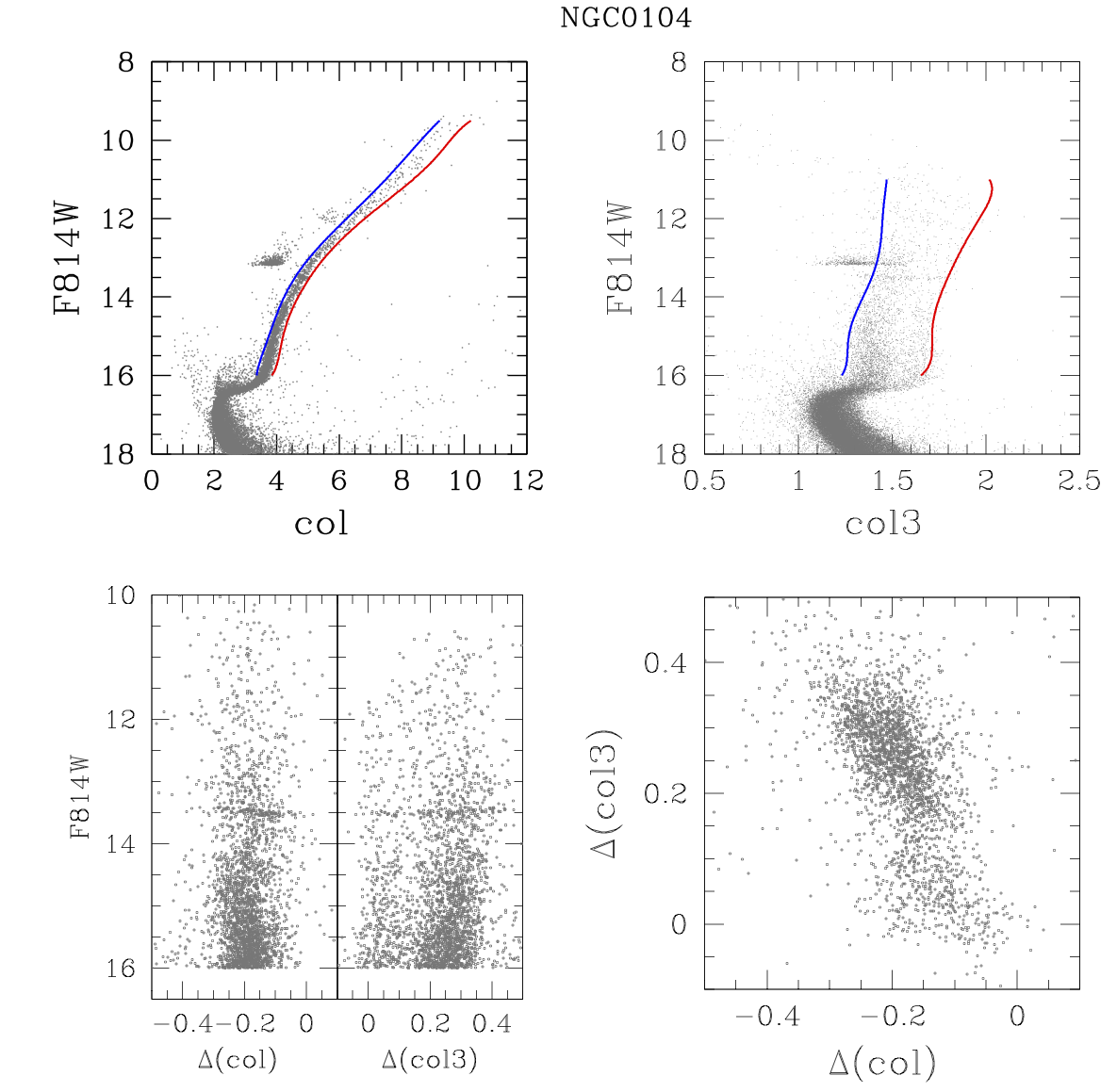}
\caption{The upper panels show the CMDs for NGC~104, $col$ versus $mag_{F814W}$
and $col3$ versus $mag_{F814W}$. The red and blue lines are the fiducials used.
Lower left panels: rectified RGBs in the normalized $col$ and $col3$. Lower
right panel: resulting PCM, where FG stars are those located below
approximately 0.1 in $\Delta col3$ and SG stars above.}
\label{f:4pannelli}
\end{figure*}

\begin{table}
\centering
\caption[]{Notation adopted in the present paper}
\begin{tabular}{l}
\hline

$col = mag_{F275W}-mag_{F814W}$ \\

$col3 = (mag_{F275W}-mag_{F336W}) - (mag_{F336W}-mag_{F438W})$ \\

 $\Delta col = W_{col} ( (col -xr)/(xr-xb) )$  \\
 
 $\Delta col3 = W_{col3} ( (col3 -yr)/(yr-yb) ) $  
(see Section 2) \\

 $W_{col}$ and $W_{col3}$ = widths of the RGB in $col$ and $col3$\\

PCM = pseudo-colour map = $\Delta col$ in abscissa and $\Delta col3$ in ordinate \\

\hline
\end{tabular}
\label{t:notation}
\end{table}

The quantity $col3$ is called a pseudo-colour because it is actually the
difference of two true colours. Formally, $col$ is an actual colour, but similarly to the
previous quantity it is used in a differential manner, as observed distances
from a blue and a red ridge line in suitable colour-magnitude diagrams (CMDs), for normalization purposes
(see below, Section 2). The resulting combination is called a pseudo-colour map,
PCM in the following. On these maps, giant stars are generally located in two
groups that are distinct along the ordinate by different values of $\Delta
col3$. 

Simulations with synthetic spectra hint that this behaviour can be mainly
attributed to variations in the N abundance (see e.g. Milone et al. 2017). Often
the two groups are rather well separated, but in some cases they seem to form
almost a continuous sequence, with no clear evidence for separation.
Interestingly, this does not seem clearly related to the cluster metallicity,
affecting the detectability of bi-metal molecular features like CN, as metal
poor GCs may show a neat separation (such as, e.g. NGC~4833) while more metal-rich
GCs (like NGC~6205) do not.

An attempt to link the photometric classification of MPs to the spectroscopic
direct method was made by Marino et al. (2019), mostly using abundances from our
extensive high resolution FLAMES survey in GCs (Carretta et al. 2009a). However,
we uncovered that a largely incomplete census of stars with abundances from
spectroscopy was identified and used by Marino et al. (2019) to test the
agreement of population tagging with the HST photometry.

We correct the situation in the present paper, using a full cross-matching of
stars in common between the HST photometry and spectroscopy for 22 GCs among 
those used in Marino et al. (2019). For each analysed GC, Table~\ref{t:numbers} 
shows the number of stars with spectroscopic abundances 
individuated on the pseudo-colour maps of Marino et al. (2019) in the second column, and how many of
them  have [Fe/H] and [Na/Fe] ratios (third and fourth columns). In column 5 we
report the number of stars we actually counted on the figures of the 
Na-$\Delta col3$ relations shown by Marino et al.   Column 6, 7, and 8 show how
many stars with spectroscopic abundance we were able to match and use in the
present work, while the references to the sources of abundance analysis are
given in the last column.
From this Table is evident that Marino et al. (2019) neglected a large number of
stars available in both the photometric and spectroscopic datasets.

\begin{table*}
\centering
\caption[]{Number of stars identified from spectroscopy in each GC}
\begin{tabular}{lcccccccl}
\hline
GC       & PCM& Fe & Na &Na-$\Delta$col3  &  Fe & Na &Na-$\Delta$col3	 & ref.spectroscopy \\
         & M19& M19& M19& M19  & us  & us &  us      &  	      \\
\hline   
NGC  104 & 10 & 10 & 10 & 10   & 17  & 17 &  17      & Carretta et al. (2009a) \\
NGC  288 & 18 & 18 & 18 & 18   & 25  & 24 &  24      & Carretta et al. (2009a) \\
NGC  362 &  9 &  9 &  9 &  9   & 15  & 15 &  15      & Carretta et al. (2013) \\
NGC 1851 & 10 & 10 & 10 & 10   & 26  & 26 &  26      & Carretta et al. (2011b) \\
NGC 2808 & 39 & 39 & 39 & 38   & 50  & 50 &  50      & Carretta (2015) \\
NGC 3201 & 22 & 22 & 22 & 22   & 28  & 28 &  28      & Carretta et al. (2009a) \\
NGC 4590 & 19 & 22 & 17 & 17   & 30  & 23 &  23      & Carretta et al. (2009a) \\
NGC 4833 & 14 & 14 & 11 & 11   & 18  & 14 &  14      & Carretta et al. (2014) \\
NGC 5904 & 10 & 10 & 10 & 10   & 16  & 16 &  16      & Carretta et al. (2009a) \\
NGC 6093 &  9 &  9 &  8 &  8   & 31  & 29 &  29      & Carretta et al. (2015) \\
NGC 6121 & 11 & 11 & 11 & 11   & 27  & 27 &  27      & Marino et al. (2008) \\
NGC 6205 & 23 & 23 & 23 & 23   & 26  & 26 &  26      & Johnson \& Pilachowski (2012) \\
NGC 6254 & 21 & 21 & 18 & 18   & 32  & 25 &  25      & Carretta et al. (2009a,b) \\
NGC 6388 &    &    &    &      & 35  & 35 &  35      & Carretta \& Bragaglia (2023) \\
NGC 6397 & 19 & 19 & 13 & 13   & 28  & 18 &  18      & Carretta et al. (2009a,b) \\
NGC 6535 & 13 & 13 & 11 & 11   & 19  & 16 &  16      & Bragaglia et al. (2017) \\
NGC 6715 & 18 & 18 & 18 & 18   & 34  & 34 &  34      & Carretta et al. (2010b) \\
NGC 6752 & 22 & 22 & 20 & 20   & 27  & 25 &  25      & Carretta et al. (2007,2009b) \\
NGC 6809 & 11 & 11 &  7 &  7   & 27  & 21 &  21      & Carretta et al. (2009a) \\
NGC 6838 & 14 & 14 & 14 & 14   & 23  & 23 &  23      & Carretta et al. (2009a,b) \\
NGC 7078 & 12 & 12 &  8 &  8   & 28  & 21 &  21      & Carretta et al. (2009a) \\
NGC 7099 & 17 & 17 &  7 &  7   & 27  & 16 &  16      & Carretta et al. (2009a,b) \\

\hline
\end{tabular}
\label{t:numbers}
\end{table*}

In this work we will focus on Na as a favoured indicator, very well suited for
tagging different stellar populations in GCs. Produced in the NeNa chain, it is
tightly linked to the O depletion (providing  the enhancement of N) in the Na-O
anti-correlation, which is  a feature found in almost every GC studied to date
and such a key feature to be assumed as a true definition of genuine GCs
(Carretta et al. 2010a). Sodium is not directly related to photometric sequences,
but is a good proxy for chemical alterations due to the MP phenomenon, maybe
even better than O.  Proton-capture reactions do enhance this species in SG
stars with respect to the plateau imprinted by supernova nucleosynthesis, so
that the abundance of Na is easier to be derived than that of other
species depleted in the same nuclear processing, such as O, for instance. Furthermore, [Na/Fe]
ratios can be reliably derived even in metal-poor GCs using the strong doublet
at 5682-56-22~\AA, securing homogeneous abundances for the vast majority of GCs,
down to the lowest metallicity objects.

Our analysis is organized in the paper as follows. In Section 2 we describe the
data and the construction of the PCM.  In Section 3 we show our first  results
on the comparison between the spectroscopic and photometric classification of MPs.
These results are discussed in Section 4, where the population ratios obtained
from different methods are tested against a sample of unpolluted, FG-like field
stars. Our present work is summarised in Section 5.

\section{The database of pseudo-colour maps (PCMs)}

The only data published in Marino et al. (2019) are average abundances of
elements from literature, largely from our FLAMES survey (Carretta et al. 2006,
Carretta et al. 2010a and references therein). In particular, most Na abundances
are from our analysis of GIRAFFE spectra in 15 GCs (Carretta et al. 2009a, see
also Table~\ref{t:numbers}). 
Unfortunately, no PCM has ever be made public by that collaboration, despite
what pledged in Piotto et al. (2015) and Marino et al. (2019). Although a PCM
is sometimes used by other groups in literature, the only public PCM we are aware
of is the one of NGC~362 published by Vargas et al. (2022). We were then forced
to construct our own PCM starting from scratch, exploiting the photometric
catalogues of HST photometry in GCs by Nardiello et al. (2018), following the
procedure illustrated e.g. in Milone et al. (2017) and Dalessandro et al.
(2019).

From the dataset of our FLAMES survey we selected 19 GCs with HST photometric
catalogues in Nardiello et al. (2018) as target GCs. To this sample we added 2
GCs with abundance analysis for a large number of stars from other sources (see
Table~\ref{t:numbers}),  for consistency and safety checks.  To this set we
added NGC~6388, whose large spectroscopic sample of 185 stars, studied in  Carretta and
Bragaglia (2023), was not available in 2019. For the 20 GCs in our survey, we
used results from Carretta et al. (2009a), and other original papers of the
series, adding data from Carretta et al. (2009b) for stars with only UVES
spectra available.

The photometric information (coordinates, magnitudes, and quality flags of
stars) comes from the HUGS database, that is the Large Legacy Treasury Program
{\em HST UV Globular Cluster Survey} (Piotto et al. 2015, Nardiello et al.
2018).

\begin{figure*}
\centering
\includegraphics[scale=0.9]{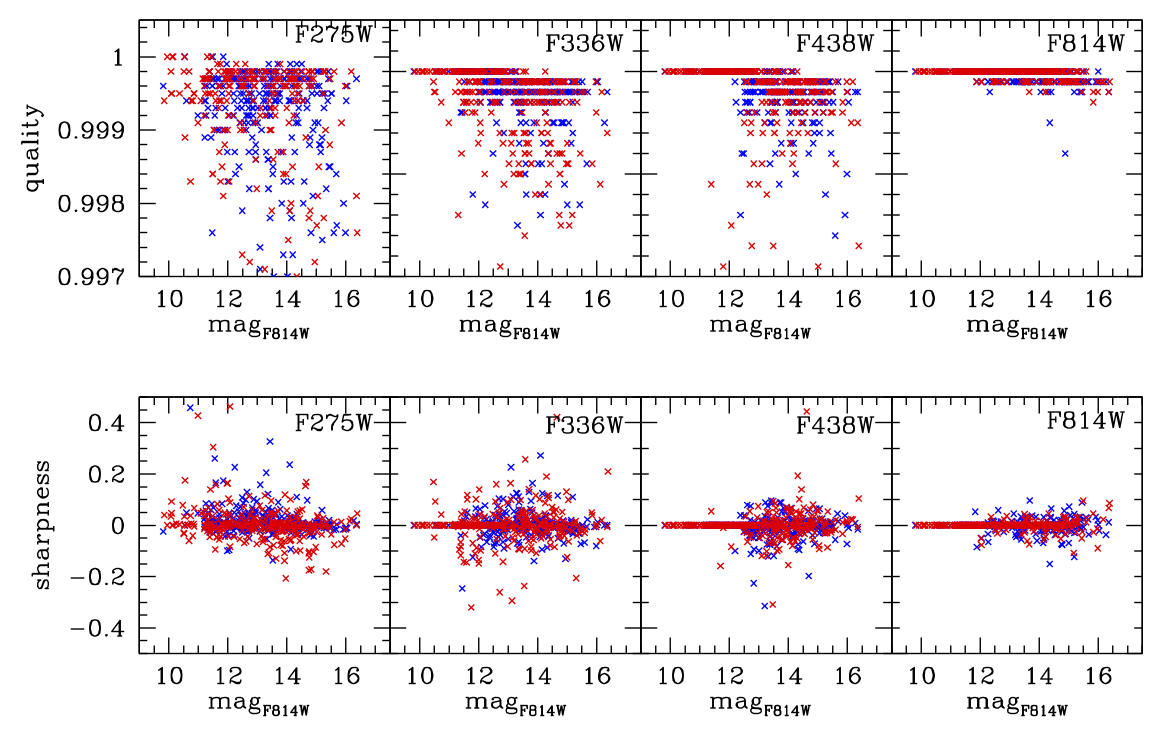}
\caption{Comparison of photometric parameters for the group of stars in all GCs 
with spectroscopy found in Marino et al. (2019; blue crosses) and those added
only in the present work (red crosses), as a function of the magnitudes in
F814W. The parameters quality and sharpness are from the catalogues of Nardiello
et al. (2018) used to construct the PCMs. }
\label{f:cfrqs}
\end{figure*}

\begin{figure}
\centering
\includegraphics[scale=0.9]{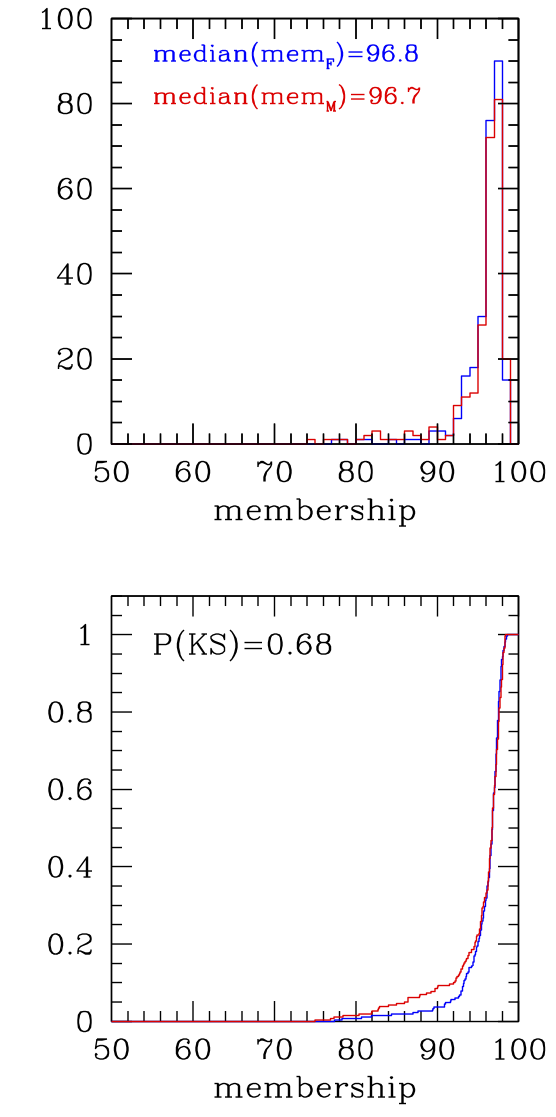}
\caption{Distribution of the membership parameter from Nardiello et al. (2018)
for stars only used by Marino et al. (2019: blue line) and only added in the
present work (red line). Upper panel: histogram of the membership parameters.
Lower panel: cumulative distribution.}
\label{f:cfrmem}
\end{figure}

\subsection{Building a PCM}

The procedure we followed is described here for one GC (NGC~104) and results for
all clusters are presented in Appendix A. We downloaded the relevant files from
the HST MAST archive\footnote{{\tt https://archive.stsci.edu/prepds/hugs/}},
choosing {\tt meth01}, the more suitable for bright stars such as the RGB ones
we are dealing with. We selected stars in the RGB region using {\sc topcat}
(Taylor 2005) and drew the blue and red fiducials for it. The lines were
computed in magnitude bins (using $mag_{F814W}$) starting about 1.5 mag brighter
than the main sequence turn-off, taking the 4th and 96th percentiles in colour
($col$ and $col3$, see Table~\ref{t:notation} for the definition) of the stars
in each bin; some manual adjustments were made where too few stars were present,
for instance above the HB level in most clusters. We then used a polynomial
interpolation to produce finer and smoother fiducial lines.
An example of the procedure followed by us is illustrated in
Fig.~\ref{f:4pannelli}, where we show these fiducials in the upper panels,  on
the colour-magnitude diagrams (CMDs) $col$ versus $mag_{F814W}$ on the left and
$col3$ versus $mag_{F814W}$ on the right.

No cut in photometric error was applied, as errors are tiny for bright
stars (less or much less than 0.01 mag). Similar considerations are valid for
other parameters such as sharpness or goodness of the PSF fits (quality),see
Fig.~\ref{f:cfrqs}.
For each of the selected star we then
computed the distance from the blue and red fiducials using eqs. 1 and 2 in
Milone et al. (2017, see also Table~\ref{t:notation}). To achieve the highest
homogeneity possible, we adopted their widths of the RGB in $col$ and in
$col3$ (see Milone et al. 2017 for details and their Table~2 for the values).

The corresponding $\Delta col$ and $\Delta col3$ versus mag$_{F814W}$ (i.e. the
verticalized RGBs) are shown in
the lower left panels of Fig.~\ref{f:4pannelli}.
Finally, the lower right panel of the figure shows the resulting PCM,
where there is a separation of stars into two main groups, the one at lower
$\Delta col3$ values representing the FG stars and the one at higher $\Delta
col3$ values indicating the SG component (see Milone et al. 2017).

In Appendix A we provide the same steps of the procedure for the remaining GCs.
We will publish at the CDS Strasbourg the whole set of PCM for the entire sample
of more than 31,800 giants in 22 GCs used in the present work, with the format
$\Delta col$, $\Delta col3$, RA, Dec, ID, where ID is a 8-digit identifier from
Nardiello et al. (2018), unique for each star in a given GC.

\subsection{Cross matching with the spectroscopic database}

As the next step, we cross-matched the spectroscopic samples and  the PCMs
derived from HST photometry (see above) using {\sc topcat} (Taylor 2005) for
each GC, retaining only  stars in common (see Table~\ref{t:numbers} for the
number of matches compared to the matches used in Marino et al. 2019).
Beside simple matches of coordinates
we used the precaution of looking at the differences between the $mag_{F606W}$
and the ground-based V magnitudes adopted in the individual papers.
Large differences indicate a probable mis-identification with a star within the
separation range adopted for the match (1 arcsec), but usually much fainter
than the spectroscopic target. In such cases, we inspected the possible
different candidate matches, choosing the one possibly slightly more distant
from the spectroscopic target, but within about 0.3-0.5 mag from its 
V magnitude, to allow for differences between the Johnson and the F606W HST
filter.

No data or star identifications were published in Marino et al. (2019). However,
from the figures shown in that paper, and knowing the Na, O abundances from the
published papers, we were able to identify all stars used in Marino et al. in
their comparison with spectroscopic results with good confidence. A few
problematic cases are reported and discussed in Appendix B. 

In Table~\ref{t:trovate} we list the stars with both spectroscopy and HST
photometry available in each of the 22 GCs of the present work. The star ID is
from Nardiello et al. (2018) and it is unique within a given GC. Using this ID
and the corresponding RA and Dec coordinates, it is easy to retrieve the stars
from spectroscopic papers. The classification in MPs for each star (pop=1 for FG
and pop=2 for SG stars) follows the spectroscopic criterion based on Na (see
below). Flag=F indicate the stars found and used by Marino et al.
(2019), identified through the position in their figures. The flag=M is used for
additional stars in common between HST catalogues by Nardiello et al. (2018) and
the papers of abundance analysis listed in Table~\ref{t:numbers}, found by
us and instead missing in Marino et al. Here we show only the excerpt of
Table~\ref{t:trovate} concerning NGC~104. The complete Table is
available at the CDS.

\begin{table}
\centering
\caption{Stars with Na abundances retrieved in HST photometric catalogues (The
complete Table is available at CDS)}
\begin{tabular}{rlrrr}
\hline
GC  &  starID  &  [Na/Fe]  & pop. & flag \\
\hline
104 & R0018794 & 0.375 & 1 & F \\
104 & R0008856 & 0.101 & 1 & F \\
104 & R0014514 & 0.388 & 1 & F \\
104 & R0008669 & 0.626 & 2 & F \\
104 & R0010831 & 0.435 & 2 & F \\
104 & R0018788 & 0.417 & 2 & F \\
104 & R0022191 & 0.504 & 2 & F \\
104 & R0006211 & 0.731 & 2 & F \\
104 & R0015269 & 0.691 & 2 & F \\
104 & R0021007 & 0.674 & 2 & F \\
104 & R0009274 & 0.686 & 2 & M \\
104 & R0011193 & 0.528 & 2 & M \\
104 & R0016453 & 0.514 & 2 & M \\
104 & R0018702 & 0.540 & 2 & M \\
104 & R0023714 & 0.619 & 2 & M \\
104 & R0024168 & 0.519 & 2 & M \\
104 & R0024181 & 0.574 & 2 & M \\

\hline
\end{tabular}
\label{t:trovate}
\end{table}

\begin{figure*}
\centering
\includegraphics[scale=0.5]{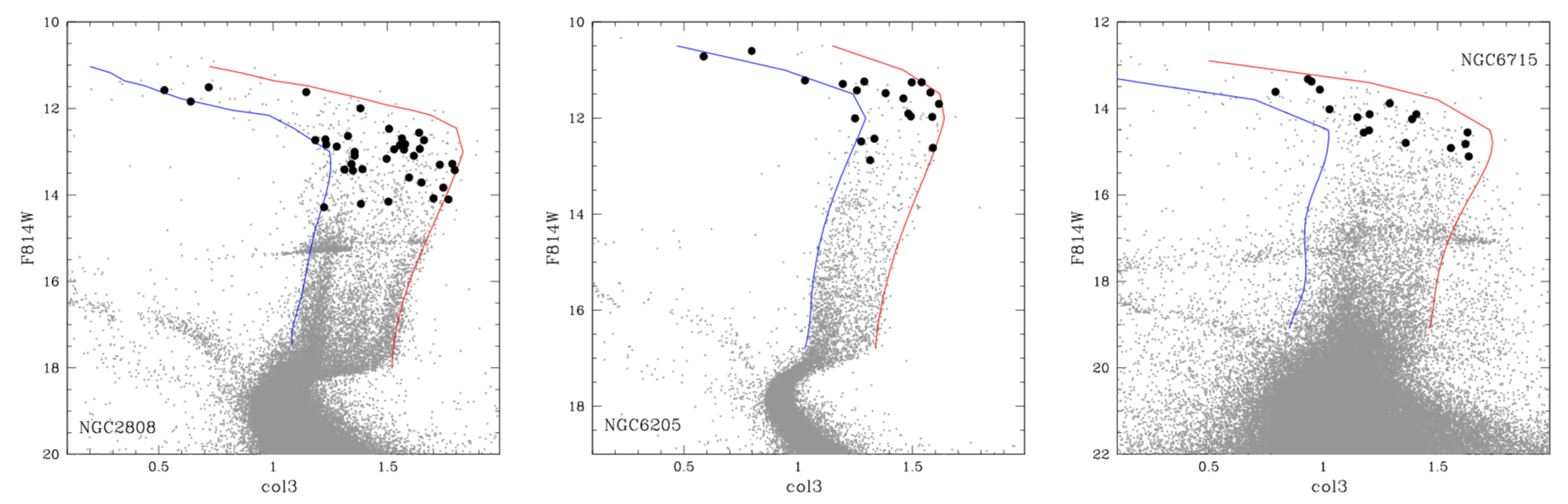}
\caption{Examples of the saturation effect in the $col3,F814W$ plot for three 
GCs in our sample. Filled black circles are stars with spectroscopy available
used by Marino et al. (2019). These stars lie also outside the vertical
distribution of the RGB, so we extended the fiducial lines to bracket also these
objects.}
\label{f:cappello}
\end{figure*}

\begin{figure*} 
\centering
\includegraphics[scale=0.7]{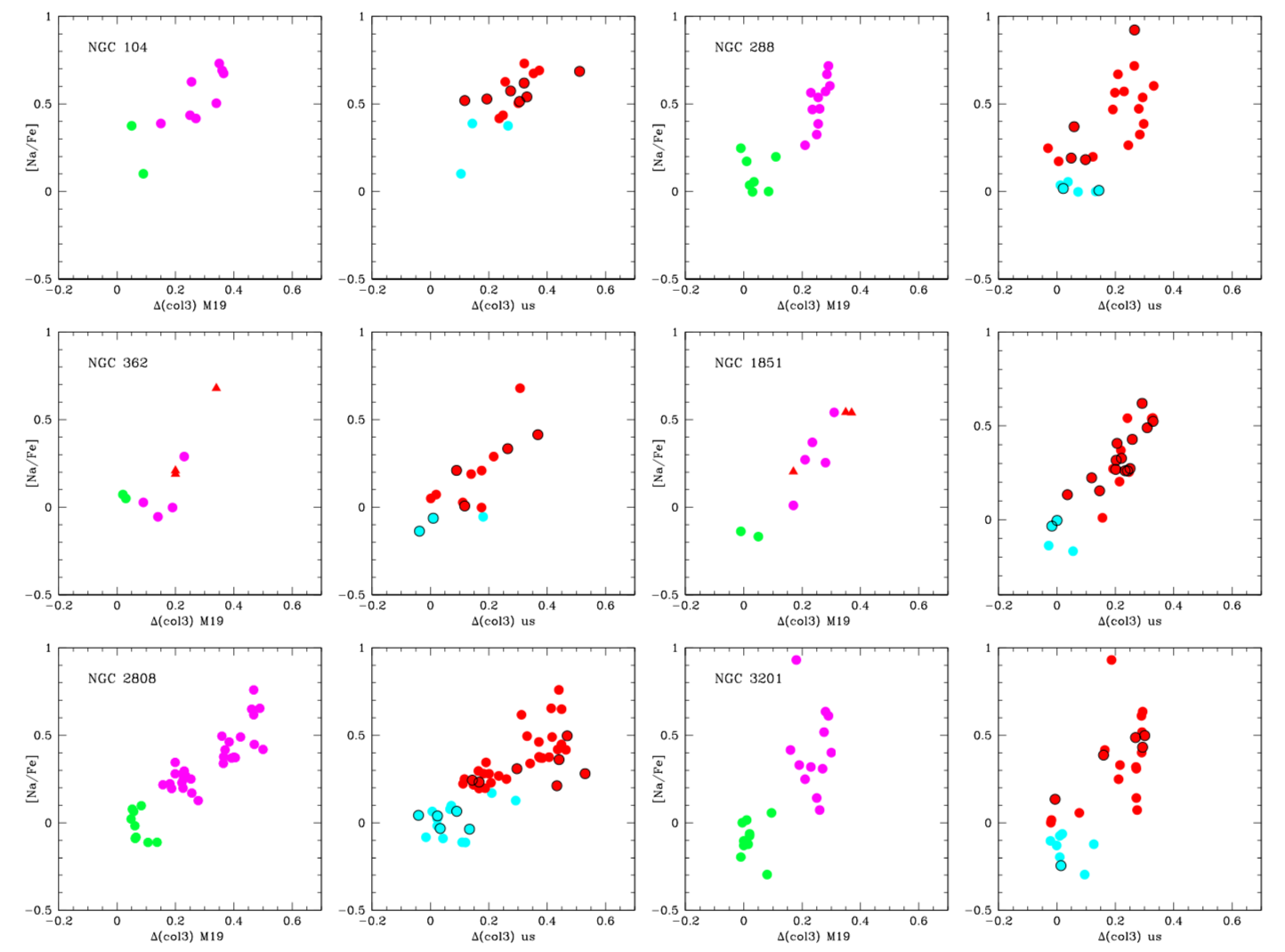}
\caption{Relations $\Delta col3$-[Na/Fe] for NGC~104, NGC~288, NGC~362,
NGC~1851, NGC~2808, and NGC~3201. For each GC, in the left panel the green and
magenta symbols are FG and SG stars, respectively, as classified from HST
photometry (through the PCMs). The red triangles represent stars in the
so-called red-RGB region. In the right panel, for each GC black circled symbols
are stars in common between photometry and spectroscopy, yet neglected in Marino
et al. Here, FG and SG stars (cyan and red, respectively) are separated
according to the spectroscopic Na abundances (see text).}
\label{f:nadc1}
\end{figure*}

\begin{figure*}  
\centering
\includegraphics[scale=0.7]{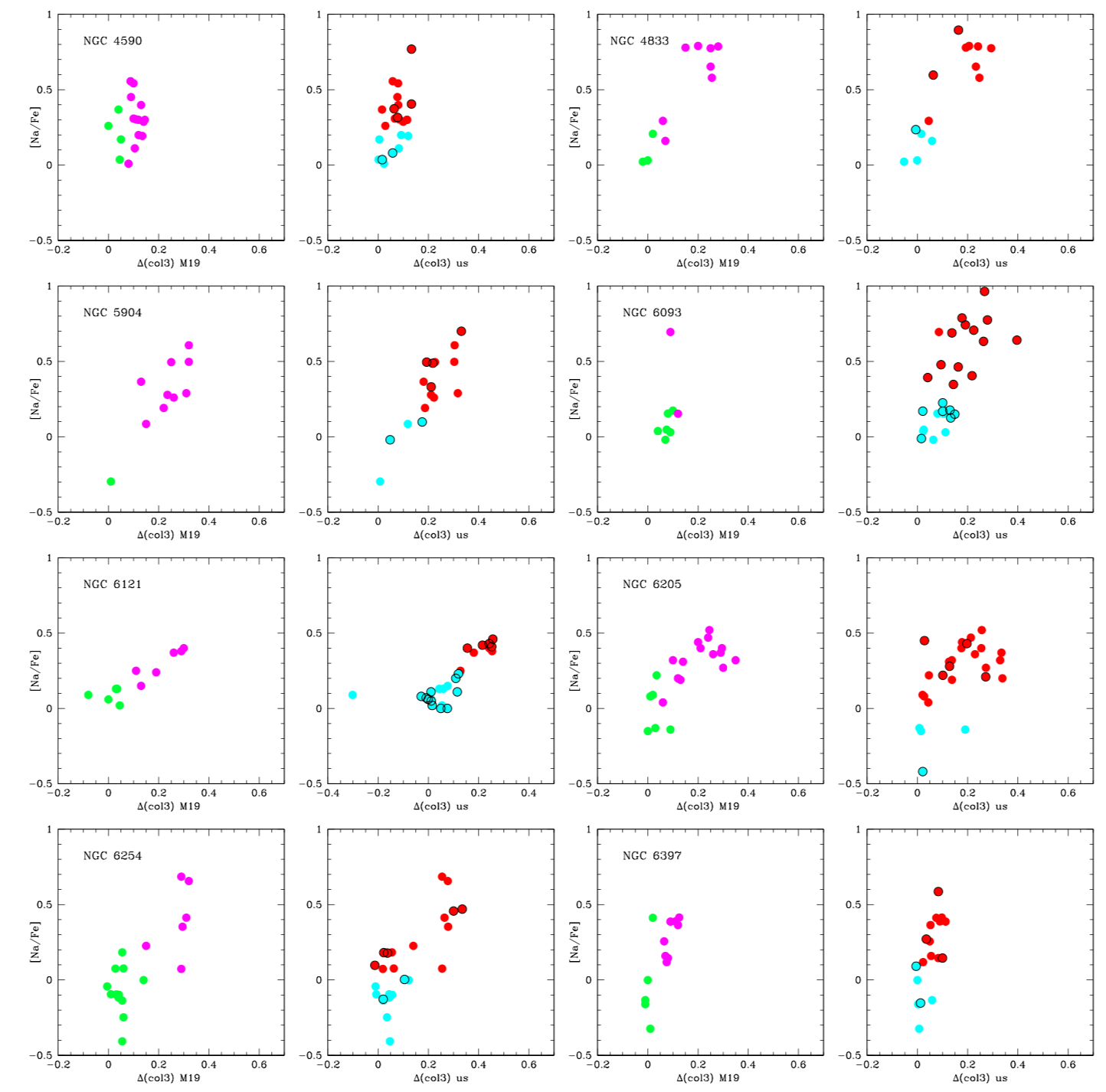}
\caption{As in Fig.~\ref{f:nadc1}, but for NGC~4590, NGC~4833, NGC~5904,
NGC~6093, NGC~6121, NGC~6205, NGC~6254, and NGC~6397}
\label{f:nadc2}
\end{figure*}

\begin{figure*}  
\centering
\includegraphics[scale=0.7]{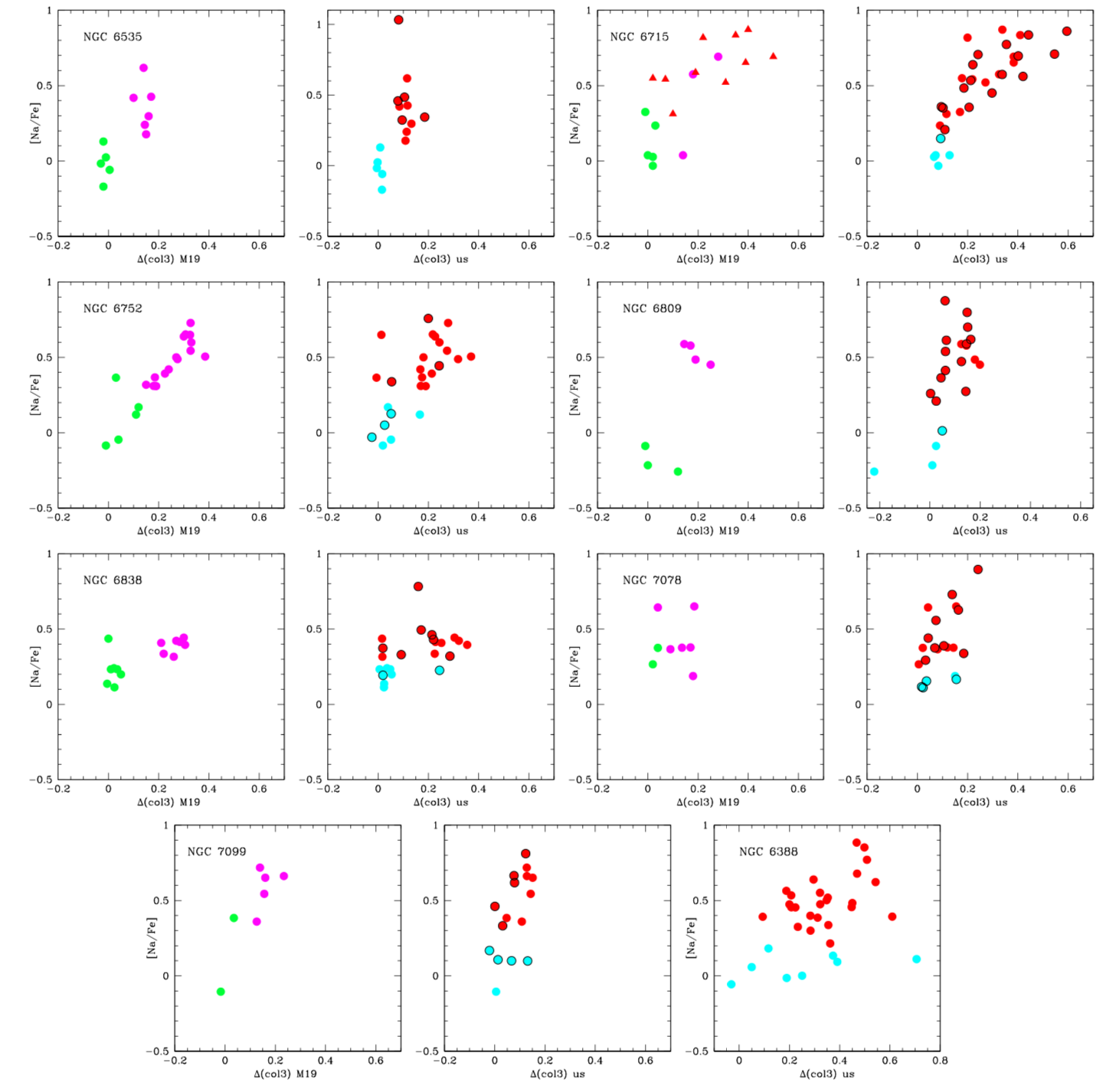}
\caption{As in Fig.~\ref{f:nadc1}, but for NGC~6535, NGC~6715, NGC~6752, 
NGC~6809, NGC~6838, NGC7078, NGC~7099, and NGC~6388. For the last GC only the relation 
from our data in the present paper is available.}
\label{f:nadc3}
\end{figure*}

As clear from Table~\ref{t:numbers}, there are many more stars with
spectroscopic abundances than those used by Marino et al. (2019), even on the
limited field of view sampled by HST observations. We were able to identify in
the 22 GCs of the present sample 298 stars used by Marino et al., all of them
with a spectroscopic Na abundance\footnote{This number is slightly different
from the total of the fifth column of Table~\ref{t:numbers} due to some
problematic stars, see Appendix B}. To this set we were able to add
291 stars (231 with Na), doubling the dataset of objects with both HST
photometry and spectroscopy available. In the following, the stars used in the
present work in addition to those adopted by Marino et al. are indicated in all
the plots as bordered in black. 

It is not clear why Marino et al. (2019) neglected a large number of
stars in each GC. To understand whether we selected stars differently from Marino et
al. (2019) we checked the distribution of a few parameters on photometric quality and
membership taken from the catalogues of Nardiello et al. (2018), used to 
construct the PCMs. In this exercise, we consider as if they were different 
samples the stars with flag=F and those with flag=M.

In Fig.~\ref{f:cfrqs} we compare the photometric parameters sharpness and
quality, as given by Nardiello et al. (2018), as a function of the
mag$_{F814W}$. Blue crosses are for the limited sample of stars with
spectroscopic abundances used in Marino et al (2019). Red crosses indicate
stars added in the present work. The two samples occupy essentially the same
region. 

The membership is compared in Fig.~\ref{f:cfrmem}. The median value for the
membership is the same for both groups, and a Kolmogorov-Smirnov test (bottom
panel) returns a probability that does not allow to reject the null hypothesis
that the two sample are extracted from the same parent population.
We further note that Marino et al. (2019) claimed that only stars that according
to their proper motion are cluster members were included in their study.
However, we uncovered that about 30 stars ($\sim 11\%$ of their adopted stars
with spectroscopy) are listed in the catalogues by Nardiello et al. (2018) with
mem=-1, meaning that the membership from proper motions is not available. 
A similar fraction ($11\%$) is found among the stars added in the present
work, increasing to about to $13\%$ when considering only stars with a Na
abundance.
Anyway, all stars with spectroscopy are cluster members according to their
radial velocities and metallicities, which is probably a more robust criterion
with respect to proper motions in particular within the crowded central regions
of the GCs. Hence we consider members all the stars flagged either F or M,
disregarding the statement by Marino et al. (2019).

\subsection{How to deal with very bright stars}

A number of stars with spectroscopic abundances are flagged with '99' in the
catalogues by Nardiello et al. (2018), indicating that the stars in question are
saturated in one or more frames. A posteriori, this is expected, since high S/N
spectra are required to obtain accurate abundances at high resolution, and
usually only the brightest stars are observed in GCs. Nevertheless, many of
these bright stars were used by Marino et al. (2019), meaning that even these
stars have a correspondence on their PCMs. Other bright stars were
identified and added in the present paper.

Using the stars identified by Marino et al. (2019) we were able to place them on
the PCM and in various diagrams, in particular the mag$_{F814w}$ vs $col3$
diagram. The brightest stars follow a sharp turn to the left in the
mag$_{F814w}$ vs  $col3$ diagram, so conservatively we in primis truncated the
RGB fiducial  lines there when deriving the PCMs (see Fig.~\ref{f:4pannelli}).
However, when we counter-identified the stars observed spectroscopically, we saw
that in half of the GCs, many were close to this limit or exceeded it, even in
the sample of stars used by Marino et al. (2019).

For those clusters (NGC~362, NGC~2808, NGC~4833, NGC~6093, NGC~6205, NGC~6388,
NGC~6715, NGC~6809, NGC~6838, NGC~7078, and NGC~7099) we then extended the
fiducial lines by eye to follow the RGB stars distribution. While not explicitly
described in  Milone et al. (2017) or Marino et al. (2019), this permitted to
place the stars in reasonable  positions in the PCMs. An example of this  is
shown in Fig.~\ref{f:cappello} for NGC~2808, NGC~6205, and NGC~6715. 
Stars of the spectroscopic sample  used in Marino et al. (2019) are over-plotted
as larger filled symbols. Several stars in the saturated region are considered
in the work of Marino et al. so that they likely used the same procedure,
extending the fiducial lines up to this region.

\subsection{A few words on the impact of reddening}

As a general rule, we did not apply correction for reddening, since this would
move all stars in the same way while the essence of PCMs is working with
differences with respect to fiducial sequences.

However, we checked that disregarding the differential
reddening present in a few GCs has no significant impact. We used NGC~3201, a
cluster well known to be affected by differential reddening, and
we applied a random variation of $\pm$0.03 in the average E(B-V) for each star.
This amount is compatible with the range of differential reddening recently
found by Cadelano et al. (2024) over the limited HST field of view centered on
this GC (about 0.06 mag peak-to-peak).

After this random addition, we repeated the complete procedure to build the new
PCM of NGC~3201.  The resulting PCM features resulted a little less compact, but
perfectly able to separate FG from SG stars on its basis. This is enough for the
goal of our paper, waiting for the publication of the precise, differential
reddening-corrected PCMs by the Legacy Program group.

\section{Spectroscopic and photometric classification of MPs}

We have now at our disposal  the newly derived PCMs, the stars with both
photometry and spectroscopy available used by Marino et al., as well as the new
additions (sometime very numerous) presented in the present paper. This means that we
have a  large database of 22 GCs where it is possible to compare the classification
of MPs based on the HST photometry (indirect way) with that based on high resolution
spectroscopy (direct way).

The comparison is shown in the next Sections.

\subsection{The Na-$\Delta col3$ relations}

As a first step to understand the agreements or possible discrepancies in the
classification of MPs with spectroscopy and photometry, we proceeded to
scrutinize the Na-$\Delta col3$ relations as a main diagnostic tool.

In the pair of panels for each GC in Fig.~\ref{f:nadc1}, on the left we 
reproduce  the relations exactly as given in Marino et al. (2019; their figures
9 and 15). The first six GCs in our sample, NGC~104, NGC~288, NGC~362, NGC~1851,
NGC~2808, and NGC~3201, are plotted in Fig.~\ref{f:nadc1}  In Fig.~\ref{f:nadc2}
and Fig.~\ref{f:nadc3} we show the other GCs in our sample. 

The green and magenta colours are adopted for FG and SG stars, respectively,
having Na (and Fe) abundances from spectroscopy. Red triangles in NGC~362,
NGC~1851, and NGC~6715 indicate stars located on the so called red-RGB sequence,
defined as an additional SG sequence above and to the red of the main SG
distribution by Milone et al. (2017).

In the left panels of each pair, the stars have both HST photometry and high
resolution spectroscopy, but the assignment to different populations was made
in Marino et al. (2019)  on a purely photometric basis, i.e. stars belong to the
FG or the SG group if they are located in the lower or upper ``blob" on the PCM,
regardless of their actual chemical composition.

To add also stars missing in Marino et al. we had no choice but to use our own
PCMs, in order to associate a [Na/Fe] value to a given $\Delta col3$ value. The
resulting relations are plotted in the right panels of each pair of the above
figures. We used the same scale as in the left panels. However, for NGC~2808 and
NGC~6121 the scale of abscissa is different, to include all matches with our
PCMs. Black bordered symbols indicated stars ignored by Marino et al. (2019)
even if they have both photometric and spectroscopic data. The  cyan and red
colours mark FG and SG stars, respectively. However, now the populations are
assigned on a pure spectroscopic criterion.

According to the definition given in Carretta et al. (2009a), we selected stars
in the FG (i.e. the primordial P component) with O, Na contents similar to field
stars of the same metallicity. Stars are assigned to this component  if their
[Na/Fe] ratios lie in the range within [Na/Fe]$_{\rm min}$ and 
[Na/Fe]$_{\rm min} +0.3$ dex, where [Na/Fe]$_{\rm min}$  is given by the lower
envelope of the Na abundance observed in each GC.
This criterion ensures that all the FG stars, with normal composition of halo
stars, are included, since 0.3 dex is about 4$\sigma$([Na/Fe]), where
$\sigma$([Na/Fe]) is the typical internal error on [Na/Fe] in our
FLAMES survey. For consistency, we applied the same criterion also to the
samples in NGC~6121 and  NGC~6205, taken from other sources, but with similar
internal errors on Na. 

This selection offers a very homogeneous approach for all GCs in the sample, and
it is based on Na abundances, that are well measurable even in the most
metal-poor GCs, and are involved in the Na-O anti-correlation, a prominent
feature of MPs in GCs. Its extension is found to be well correlated to
the global GC mass (Carretta et al. 2010a, their fig. 15) which is one of the
main driver of the MPs phenomenon, as inferred in a number of studies (Carretta
2006, Carretta et al. 2009a,b,  Carretta et al. 2010a, Pancino et al. 2017,
Milone et al. 2017, M\'esz\'aros et al. 2020). Moreover, the choice of the
minimum Na content in GCs of different metallicity allows to follow the pattern
of field halo stars derived from the plain Galactic chemical evolution. The
soundness of the above criterion is proven by the good match between the FG
fraction and the floor of unpolluted field stars over the whole metallicity
range spanned by Galactic GCs (see Carretta 2016 and Section 4 in the present
paper).

The relations shown by Marino et al. generally present a clean separation between FG
and SG stars, with a sort of monotonic progression of Na abundances increasing
as the $\Delta col3$ values increase. Often there is a even a gap separating 
the two populations in this plane. This is simply a reflection of the adopted
photometric criterion. On the PCM, the density thickening of stars corresponding
to FG and SG stars shows only a slight superposition along the $\Delta col3$
coordinate, and this explains the aspect of these relations. According to Figure
27 in Marino et al. (2019), the spread along the $\Delta col3$ ordinate in the
PCM is measuring essentially the variation in [N/Fe]. In proton-capture 
reactions in H-burning at high temperature both N and Na are enhanced (see e.g.
Langer et al. 1993), in the ON cycle and NaNa chain, respectively, explaining
the  observed correlation.

Nonetheless,  already at a glance the photometric classification of MPs shows 
several inconsistencies, visible in the Na-$\Delta col3$ relations by 
Marino et al. (2019). Some
FG stars (e.g. in NGC~6397, NGC~6752, NGC~6838) have large Na abundances that
are different from the low values typical of the primordial population. On the
contrary, some SG stars show low Na abundances, consistent with the normal,
unpolluted stellar population in GCs. A few objects are at the border between FG
and SG stars, but in other cases (e.g. NGC~6093, NGC~6205, and NGC~6715) their
Na abundance is clearly discrepant with respect to their photometric
classification. Finally, we note that in some very metal-poor GCs (e.g.
NGC~4590, NGC~7078) FG and SG stars are totally confused, and both classes share
the same range of Na abundances.

The monotonic progression in the Na-$\Delta col3$ trend weakens when the tagging
is done using the spectroscopic criterion, as in the right panels of each pair 
in Figures from 5 to 7. Despite a smooth increase of Na along the Na-O
anti-correlation, often we do not observe a clean, almost one to one
correspondence. Instead, to a given $\Delta col3$ value corresponds a large
range of Na abundances, reaching even 0.7-1 dex in some cases.

Finally, in Fig.~\ref{f:nadc3} we show our result for NGC~6388. Marino et
al. (2019) did not consider this GC, so there is no relation Na-$\Delta col3$
from their work. We exploited the large spectroscopic sample recently assembled
in Carretta and Bragaglia (2023) and the relation is shown in the last panel of
Fig.~\ref{f:nadc3} using the usual spectroscopic tagging, as in previous
figures.

We conclude our analysis by noting that in NGC~2808 we see a FG star on the PCM
of Marino et al. (2019) which then disappears in their Na-$\Delta col3$
relation even if all the matches for NGC~2808 have a Na abundance. It was then 
impossible for us to identify this star. On the contrary, a few stars with
spectroscopy used by Marino et al. (2019) in some GCs should not appear on the
PCM, because they do not have all the required magnitudes in the catalogues by
Nardiello et al. (2018). These and a few other peculiar stars are discussed
in Appendix B.

\begin{figure*}
\centering
\includegraphics[scale=0.9]{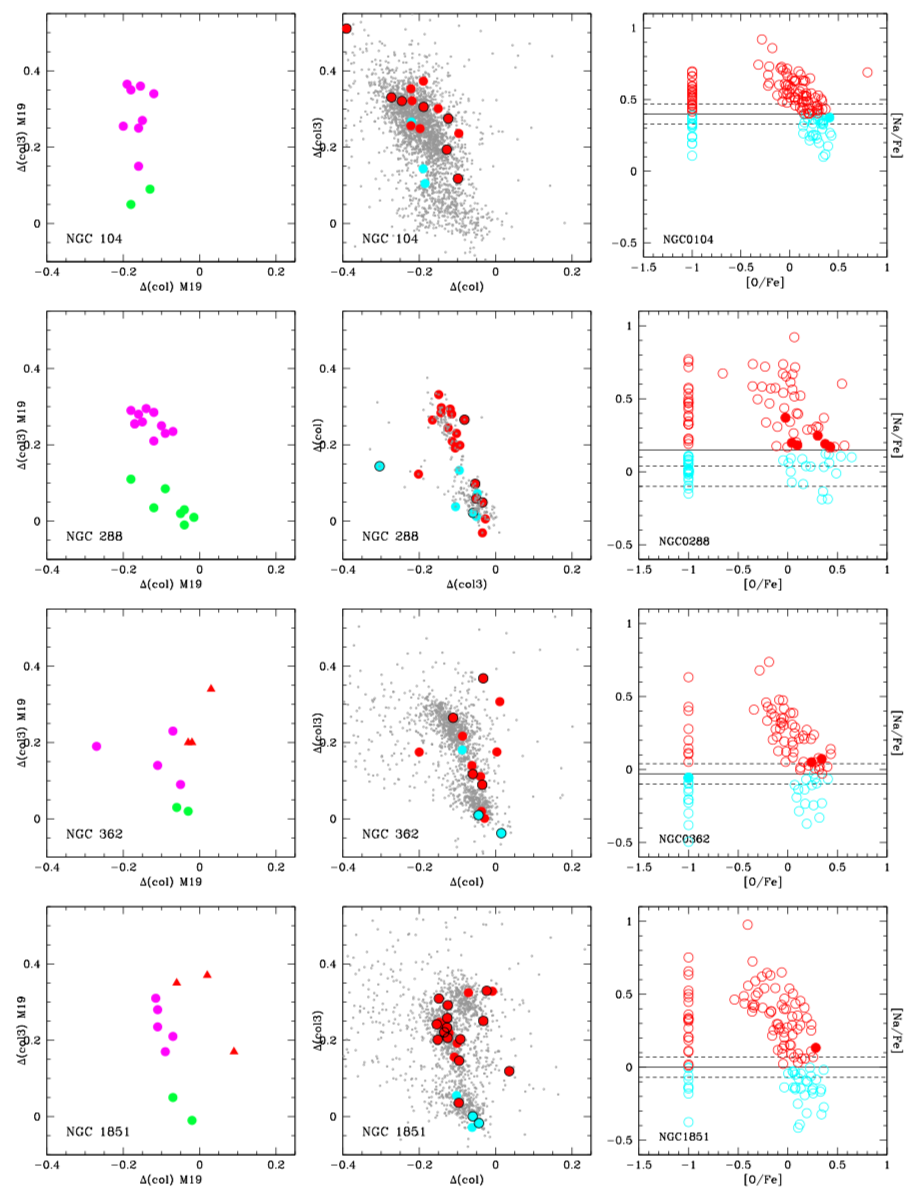}
\caption{Stars with spectroscopy on the PCMs of NGC~0104, NGC~288, NGC~362, and NGC~1851.
In the left panels the photometric classification of MPs is adopted and in the
middle panels the population assignment is based on spectroscopy. Our PCMs are
indicated with small grey dots. In the right panels, stars considered mismatches
in classification are superimposed to the Na-O anti-correlation (as filled symbols).
Solid lines in each GC indicate the limit  [Na/Fe]$_{\rm min} +0.3$ dex (with a
range $\pm 0.07$ dex, dashed lines) used to separate FG and SG stars. The colour
coding is as in Fig.~\ref{f:nadc1}.}
\label{f:chm1}
\end{figure*}

\begin{figure*}
\centering
\includegraphics[scale=0.9]{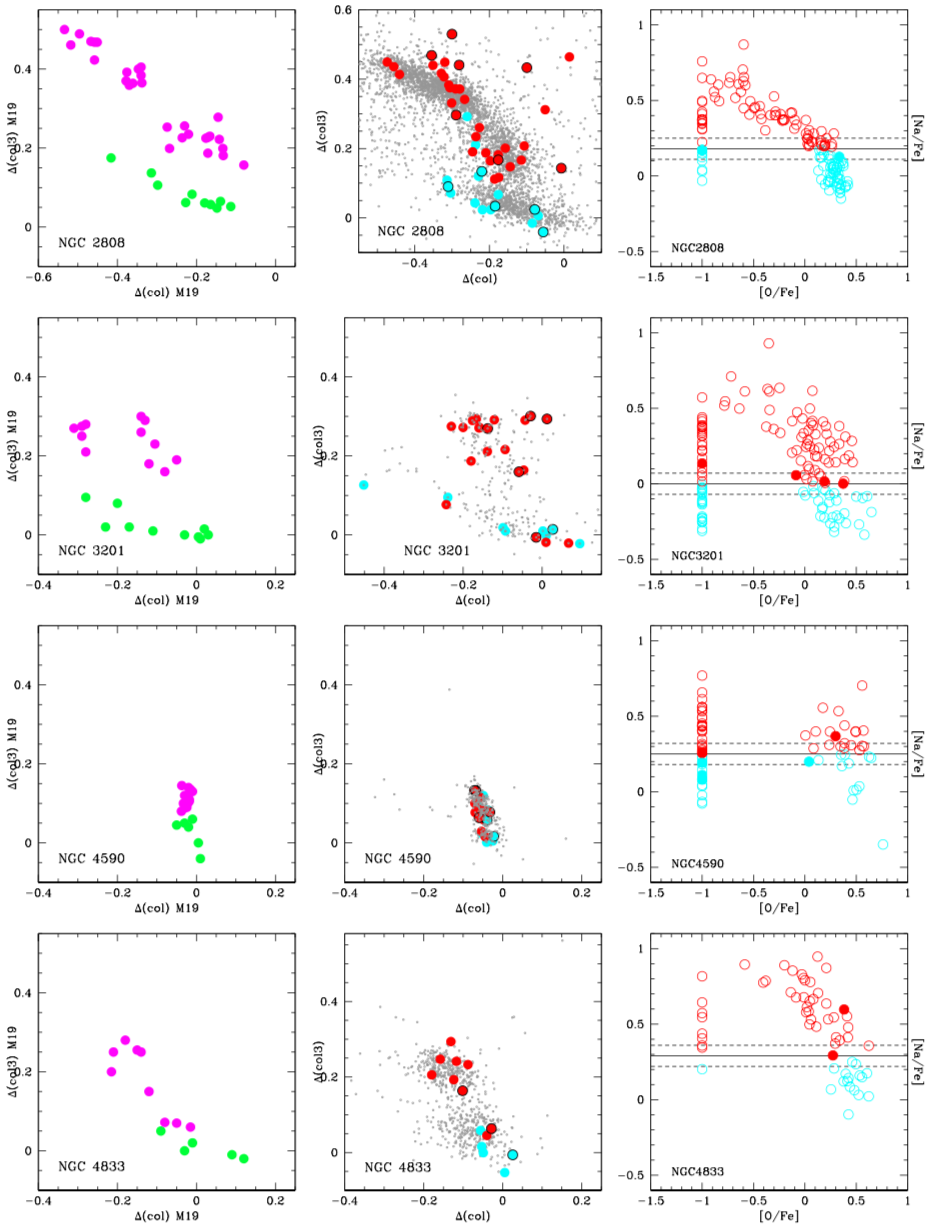}
\caption{As in Fig.~\ref{f:chm1} for NGC~2808, NGC~3201, NGC~4590, and NGC~4833.}
\label{f:chm2}
\end{figure*}

\begin{figure*}
\centering
\includegraphics[scale=0.9]{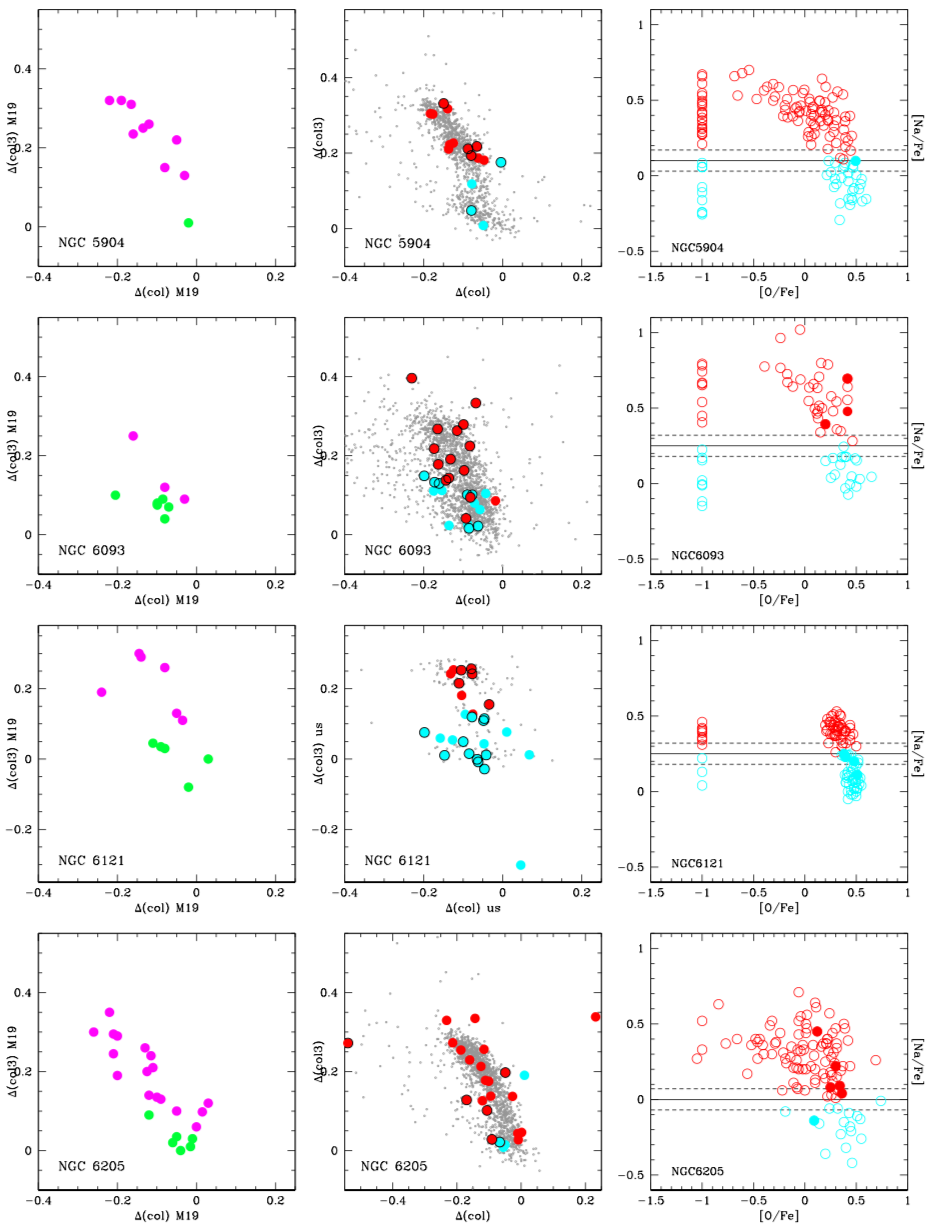}
\caption{As in Fig.~\ref{f:chm1} for NGC~5904, NGC~6093, NGC~6121, and NGC~6205.}
\label{f:chm3}
\end{figure*}

\begin{figure*}
\centering
\includegraphics[scale=0.9]{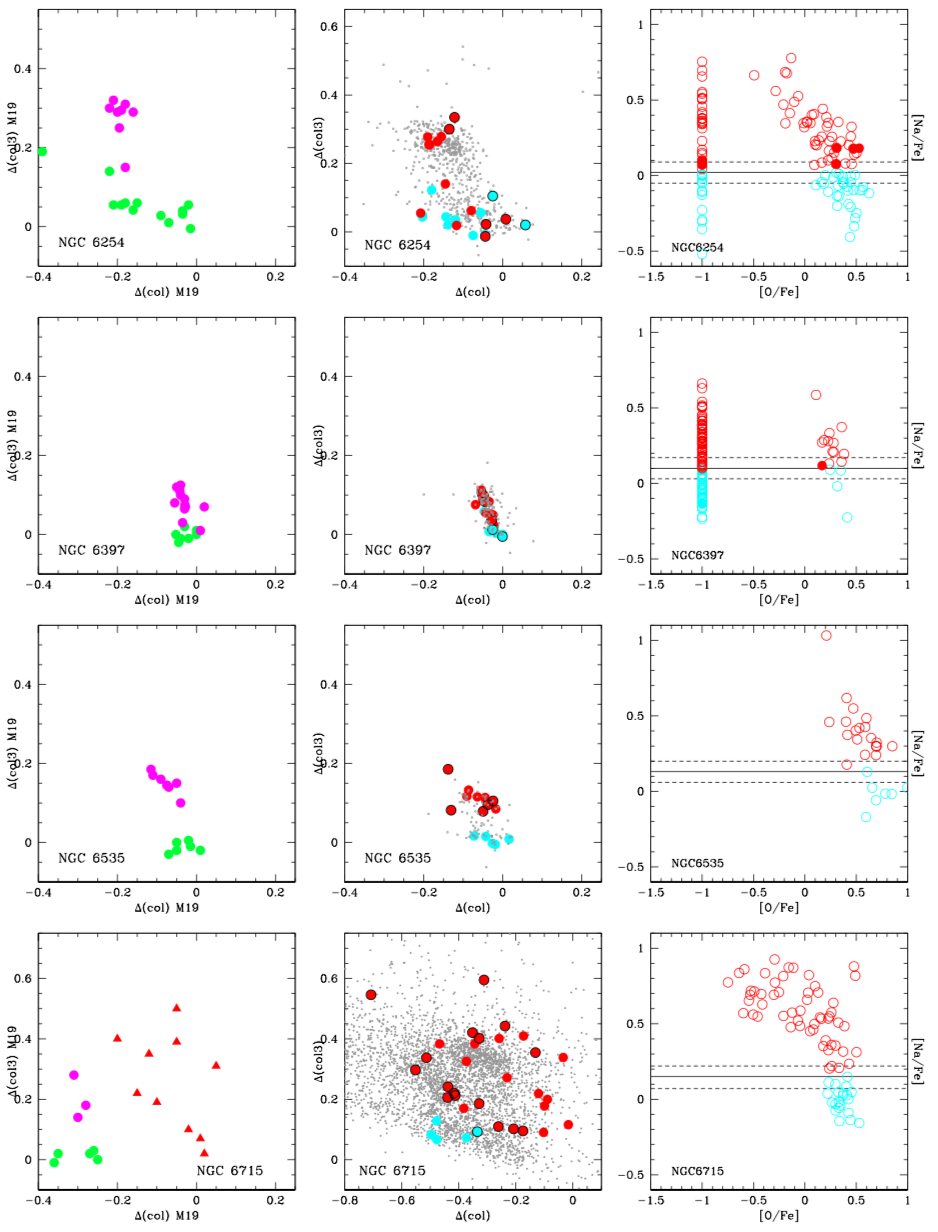}
\caption{As in Fig.~\ref{f:chm1} for NGC~6254, NGC~6397, NGC~6535, and NGC~6715.}
\label{f:chm4}
\end{figure*}

\begin{figure*}
\centering
\includegraphics[scale=0.9]{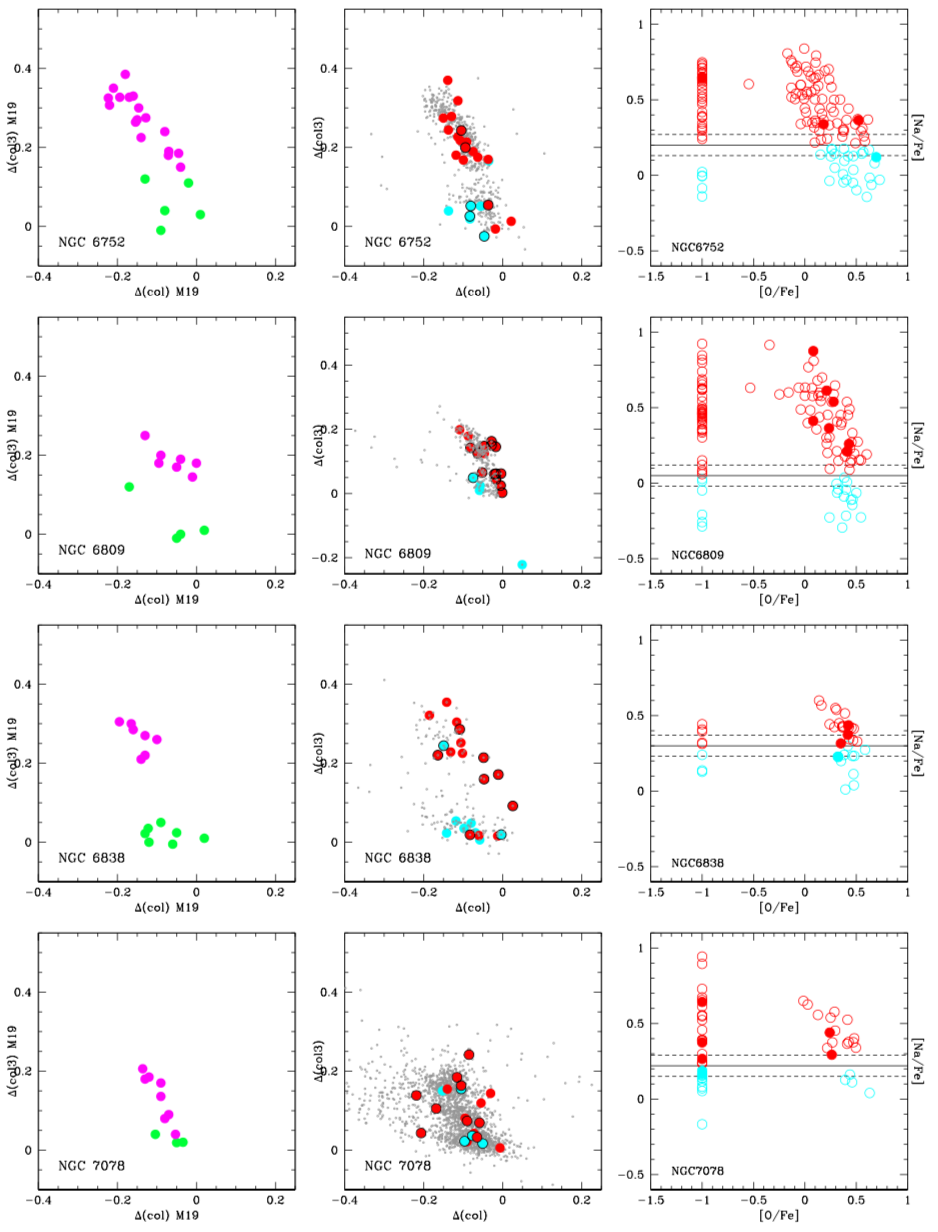}
\caption{As in Fig.~\ref{f:chm1} for NGC~6752, NGC~6809, NGC~6838, and NGC~7078.}
\label{f:chm5}
\end{figure*}

\begin{figure*}
\centering
\includegraphics[scale=0.9]{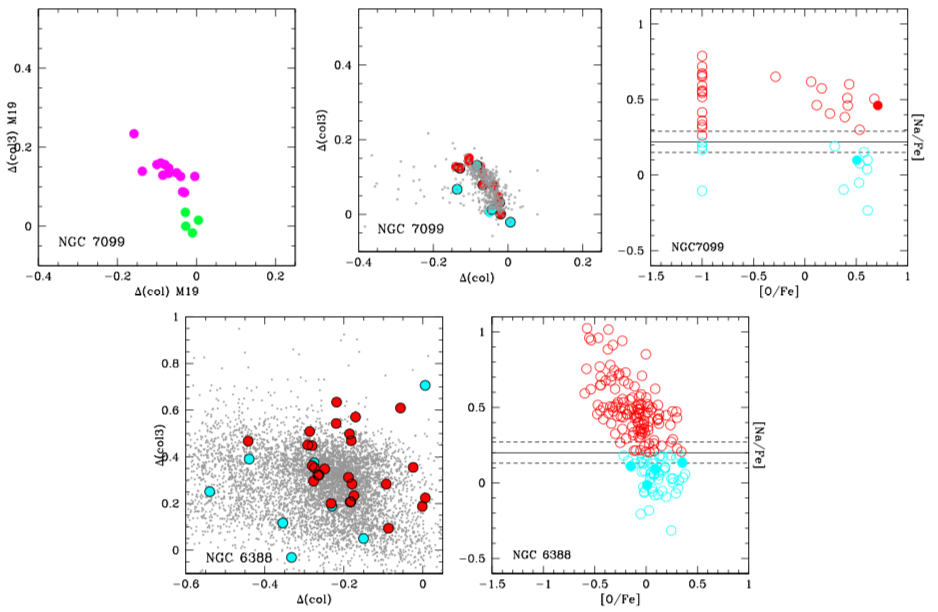}
\caption{As in Fig.~\ref{f:chm1} for NGC~7099 and NGC~6388.}
\label{f:chm6}
\end{figure*}

\subsection{Population classification on the PCMs}

The enlarged sample of stars with spectroscopic abundances and our new PCMs
available for the cross-identification of stars allow us to check in detail the
agreement or differences arising from the classification of MPs using 
photometry or spectroscopy.

The comparison is shown for the first four GCs in Fig.~\ref{f:chm1}. 
Figures from Fig.~\ref{f:chm2} to Fig.~\ref{f:chm6} are for the remainder of GCs.
On the left panels we reproduce the position on the PCMs of the stars with
spectroscopic abundances used by Marino et al. (2019). Again, green and magenta
colours are employed according to their photometric classification,
disregarding direct measures of the chemical composition of stars. 
Adopting this classification, the FG stars generally have values of $\Delta
col3$ lower than those of SG stars, likely due to a lower N abundances in their
chemical composition. However, this general rule presents a few exceptions in 
about a third of the sample of GCs (e.g. NGC~2808, NGC~6093, NGC~6205, NGC~6254,
NGC~6397, NGC~7078). In these cases, FG and SG stars are found at the same level
in $\Delta col3$.

In the central panels we plot the stars with spectroscopic abundances,
superimposed to our PCMs (small grey dots). The colour code for stars is the
same as in Fig.~\ref{f:nadc1} and follows the same assignment to MPs based on
the [Na/Fe] abundance ratio as described in Section 3.1 and visualised in the
panels on the right side for each GC. A solid line illustrates the criterion 
[Na/Fe]$_{\rm min} +0.3$ dex used to separate FG stars (cyan open circles) from
SG stars (red open circles). Dashed lines indicate the range $\pm 0.07$ dex
around the separation, corresponding to the typical internal error on [Na/Fe] in
abundance analyses from high resolution spectra (e.g. Carretta et al. 2009a).

The spectroscopic classification of MPs is made ignoring a priori where the
stars are located on the PCM, being based only on their Na content. This second
approach immediately reveals a number of mismatches, with respect to the
population division of MPs based on photometry and, ultimately, on the effect of N
abundances. Some stars belonging to the SG according to their enhanced Na are
found on the PCM in the lower group that should be populated by unpolluted stars
with the normal composition of halo stars of similar metallicity.

Vice versa, FG stars, that should display only the abundance pattern imprinted
by supernovae nucleosynthesis, are found well inside the upper group on the
PCMs, the one populated by giants with high or very high N abundances. 

The observed mismatches in classification fall in two categories. We call major
mismatches those whose Na abundance is unambiguously distant  from the
dividing line of FG and SG stars (the separation of P and I components in the
scheme defined in Carretta et al. 2009a). We define minor mismatches those whose Na
abundance lies within $\pm 0.07$ dex from the separation. In principle, this
second class of misclassifications could be due to the star to star errors
associated to the Na abundances. However, we note that the adopted criterion
for the selection of FG stars ([Na/Fe]$_{\rm min} +0.3$ dex) already includes
a generous range of more than 4 times this internal error, so that it is
unlikely that some FG stars are missing. Moreover, the internal error in Na is by
definition a random one, so we expect that FG stars
near the separation are shifted upward or SG stars are pushed downward simply
from the associated error with the same probability, so that statistically the neat effect is roughly
unchanged. 

On the other hand, a similar effect is present even in the photometric division
of MPs, where a separation line between the two groups on the PCM is adopted
with the same slope for all GCs, regardless of e.g. crowding, metallicity or
global GC mass\footnote{This assumption clashes with the knowledge that the Na-O
anti-correlation, albeit being universal among GCs, is different in extent and
shape from cluster to cluster.}.
For the photometry of bright giants in GCs the associated errors in magnitudes
are usually tiny, however  the resulting observational error on $\Delta col3$
(estimated from Fig. 2 of Milone et al. 2017 or Fig. 2 of Marino et al. 2019) 
amounts on average to about 0.05 mag. The meaning is that a similar drifting
effect is likely to impact also the photometric fraction  of MPs in GCs.
Actually, in several GCs there are a few stars that it is impossible to flag as
discrepancies or agreements between photometry and spectroscopy simply because
it is unclear their belonging to one or the other of the two groups in  the PCM,
falling equally far from both groups. To conclude, there could be a grey zone in
both methods where observational uncertainties may slightly blur the
adopted criteria for population tagging. The majority of stars is nevertheless
clearly distinct in both the photometric and spectroscopic approaches, allowing
an useful comparison of their merits and drawbacks. 

The mismatches in classification identified in the present sample of GCs are
indicated on the Na-O anti-correlation (Fig.~\ref{f:chm1} to Fig.~\ref{f:chm6})
as filled symbols, coloured red or cyan  according to their status of SG or FG
stars following the Na criterion. We list in Table~\ref{t:tabmismatches} all
stars whose classification differs between the HST photometry and the high
resolution spectroscopy. Using the  information reported in this Table it is
possible to situate each star on both our PCMs and on the Na-O anti-correlation.
In Table~\ref{t:tabmismatches} we list the value of  [Na/Fe]$_{\rm
min} +0.3$ dex and all the mismatches we found for each GC, with star ID, $\Delta col$ and
$\Delta col3$,  the photometric classification based on PCMs, the [Na/Fe] and
[O/Fe] ratios. Finally,  a flag is used to label  major mismatches with respect
to the photometric population (capital YES). The lower case flag yes warns that
the mismatch is a minor one.

Our enlarged sample in 22 GCs adds up 529 stars with both a Na abundance and a
pair of coordinates on the PCM. We found 75 mismatches in the 
population tagging on the whole, i.e. a fraction of $14\pm 2\%$ (Poissonian error). Among
these, 53 are major mismatches and only 22 minor mismatches, in the sense
explained above. We are confident that even using the limit in Na as a strict
condition is justified, and in the following we consider both  categories
together, except for noting if some results are affected by a particular
prevalence by one of the types.

\section{Discussion and conclusions}

We quantify all the observed mismatches in the sample of GCs in 
Table~\ref{t:totmismatch} where  we counted both occurrences, i.e. SG stars from
spectroscopy mimicking FG stars (according to HST photometry) and FG stars
(according to their Na abundance) posing as SG stars on the PCMs. In
Table~\ref{t:totmismatch} we report in parenthesis the fraction of these events
in each GC, computed over the total number of stars with abundance of Na found
in that GC. On average, over 20 GCs, the fraction of inconsistencies
between the two methods of tagging is about $16\%$, ranging from a minimum of
$\sim 4\%$ (NGC~1851, NGC~2808, NGC~6121) to a rather impressive maximum of
about $33\%$ of stars in NGC~6809 and NGC~7078. The
result does not depend much on the category of mismatch, being the average
fraction about $14.5\%$ if we limit to the major mismatches.

Despite the small numbers involved, the results summarized in
Table~\ref{t:totmismatch} are informative. At first glance it is possible to
infer that the largest number of discrepancies between spectroscopy and HST
photometry occurs among the most metal-poor GCs.

\begin{figure*}
\centering
\includegraphics[scale=0.95]{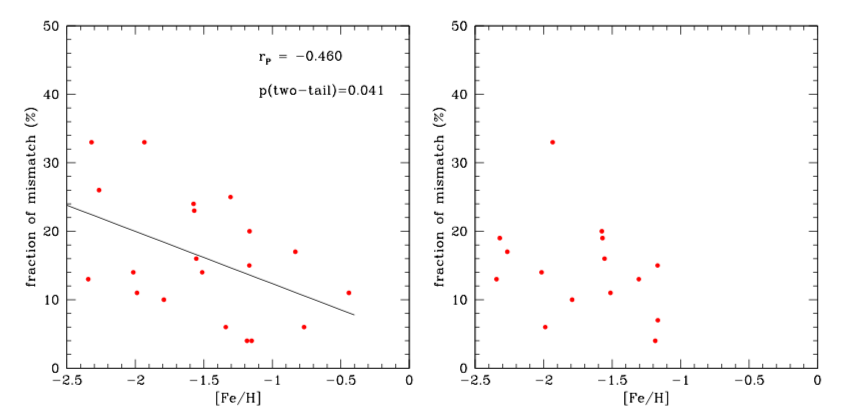}
\caption{Relation between the fraction of mismatches in population tagging and
the cluster metallicity. On the left panel all the mismatches in
Table~\ref{t:totmismatch} are counted, and the solid line is a linear fit to the
data (Pearson correlation coefficient and the two-tail probability are labelled
in the panel). In the left panel fractions are computed using only major mismatches.}
\label{f:mismatch}
\end{figure*}

This impression is visually confirmed in the left panel of Fig.~\ref{f:mismatch}
where the fraction of mismatches in the classification of MPs is plotted as a
function of the metallicity of GCs. The [Fe/H] values are from our homogeneous
metallicity scale based on high-resolution UVES spectra (Carretta et al. 2009c),
with the exception of NGC~6205, taken from Johnson and Pilachowski (2012).

There is a clear anti-correlation between the fraction of cases with a
discrepant  tagging of MPs and the cluster metallicity. Admittedly, the
Poissonian errors associated to the fractions are rather large (about 8\% on
average), due to the limited number of cross-identifications between ground-based
spectroscopy and the small field of view of HST observations, centered on the
GC centres. However, the relation appears to be statistically significant. A
linear regression returns a high value of the Pearson correlation coefficient
with a significant two-tail probability that the relation is not due to a chance
effect. Both values are indicated in Fig.~\ref{f:mismatch}, where the solid line
indicates the regression. This Figure visualizes that the fraction of stars
misclassified on the PCMs, according to their Na
abundance, is generally low in metal-rich or moderately metal-rich GCs but it
raises significantly among metal-poor GCs.

In the right panel of Fig.~\ref{f:mismatch} the fraction is computed using only
the major mismatches. We must duly note that the three most metal-rich GCs
disappear, as also the statistical significance of the relation. Nevertheless,
the trend still exists.
 
The increase in the number of discrepancies observed in more metal-poor GCs is
compatible with the reduced response of RGB colours to the strength of the
molecular bands, in particular the NH and CN features driving the behaviour of the
$\Delta col3$ coordinate on the PCM. The metallicity dependence of the
band strength for bi-metal molecules varies quadratically with the
metal abundance, so that the tagging with the photometric classification is
expected to be more fraught with uncertainties when the metallicity is
decreasing (see e.g. the discussion in Lee 2023 about the increasing confusion
in population tagging in some photometric systems for more metal poor GCs).

In general, however, the relevant number of mismatches found between photometry and
spectroscopy is a bit surprising. The N abundance is the main source of the gross
differences between FG and SG stars on the PCM, as this element directly affects
the fluxes in the bands used to build the pseudo-colours and the
resulting maps. The Na abundances do not directly affect the HST fluxes, but the
correlation between this species and $\Delta col3$ is a clear proxy of a
correlation between Na and N. 

A progressive enhancement of N along the RGB
is expected in the normal evolution of low mass stars, following the first
dredge-up and another mixing episode after the luminosity bump level on the RGB
(e.g. Charbonnel 1994; Gratton et al. 2000). However, the almost constant values
of $\Delta col3$ as a function of luminosity along the RGB (lower left panels in
the Figures of Appendix A) excludes that the observed pattern on the PCMs is
related to any mixing episode. On the contrary, the variations in N must be
intrinsic to the chemical composition of the whole star (Cohen et al. 2002)
originated by the incomplete CN and ON burning at the same temperature where the
NeNa cycle is active. 

As a consequence, the failure of PCMs to correctly locate stars according to
their Na abundance in a non negligible number of cases is at first glance a
puzzle. 
The solution to this apparent riddle is that we uncovered the modern
manifestation of the decoupling between N and heavier atomic species such as Na,
well known since the first studies on MPs (Smith et al. 2013, Smith 2015).
Extensive works by Graeme Smith and collaborators unveiled that there is often a
decoupling between the N abundances derived from molecular bands  on one hand
and the abundances of heavier elements such as O and Na on the other. This
occurrence may be indicative of a complex enrichment by polluters with a range
of masses and formation times within each GC.

\begin{figure}
\centering
\includegraphics[scale=0.7]{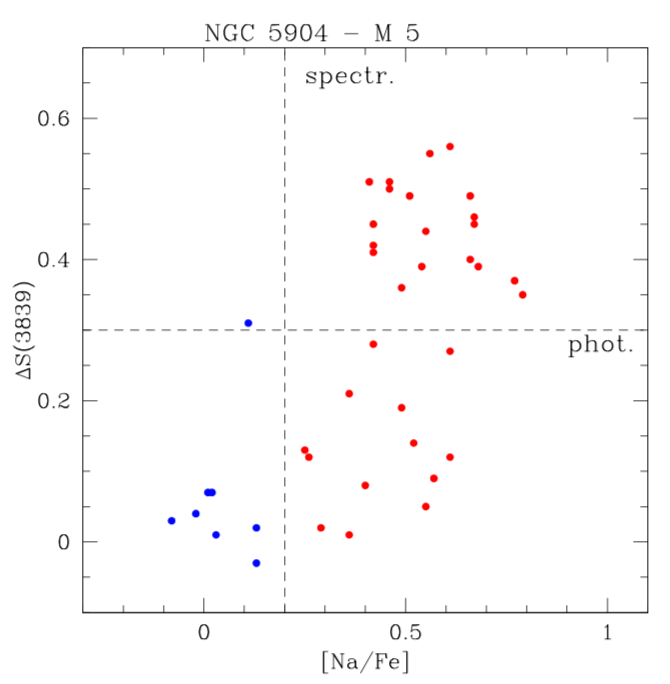}
\caption{Comparison of CN (Smith et al. 2013) and Na (Carretta et al. 2009a,b)
indicators of MP tagging in NGC~5904 for stars in common between the two 
studies. The horizontal dashed line separates CN-strong (top) and CN-weak
(bottom) stars, whereas blue and red filled circles represent FG and SG stars,
respectively, as tagged by our spectroscopic criterion based on Na abundances
(see Section 3.1).}
\label{f:cpap5}
\end{figure}

Accurate comparison of C, N and O, Na abundances in GC stars is neither easy nor
immediate, since these elements are usually measured with different techniques,
mostly low resolution spectroscopy for the former pair and high resolution
spectroscopy for the second one. 
The comparison shown in Fig.\ref{f:cpap5} represents a typical case.

The CN index S(3839) for giants in NGC~5904 (Smith et al. 2013) measuring the
$\lambda3883$ CN band strength is compared to Na abundances for RGB stars in
common from Carretta et al. (2009a,b). To compensate at first order for
different effective temperature of RGB stars, the residuals with respect to the
baseline defined by CN-weak stars as a function of magnitude, $\Delta S(3839)$,
are used. The dashed horizontal line separates CN-weak and CN-strong stars,
according to Smith et al. (2013). This level was labelled ``phot" because the
criterion dividing FG, CN-weak stars from SG, CN-strong stars is basically the
same tracing mainly the N abundance derived by the incidence of molecular bands
on photometric fluxes. The vertical line along the Na axis is labelled ``spectr"
because the division is made using the criterion defined in Section 3.1 for the
tagging of MPs based on Na abundances.

In Fig.~\ref{f:cpap5} the same evidence as given by the analysis
of PCMs from HST photometry is reproduced. There is clearly a decoupling between the
abundances of the lighter and the heavier proton-capture elements, superimposed
to a global trend, with SG stars being more enhanced in CN and Na than FG,
N-poor, Na-poor (and C-rich) stars. Even when N abundances are estimated from
low resolution spectroscopy, there are a few mismatches between the two
classifications. A number of stars assigned to the FG according to their CN
index have Na abundances as high as those classified as SG stars by both
criteria. Conversely, in the pool of objects in common one CN-strong star has [Na/Fe]
compatible with the primordial level in NGC~5904. 
The effect is striking because $^{23}$Na is produced through the
destruction of $^{22}$Ne already at 25 MK, whereas O achieve simultaneously a
depletion. However, at the above temperatures the CN isotopes have already
reached their equilibrium value (Prantzos et al. 2017)

\begin{figure*}
\centering
\includegraphics[scale=0.7]{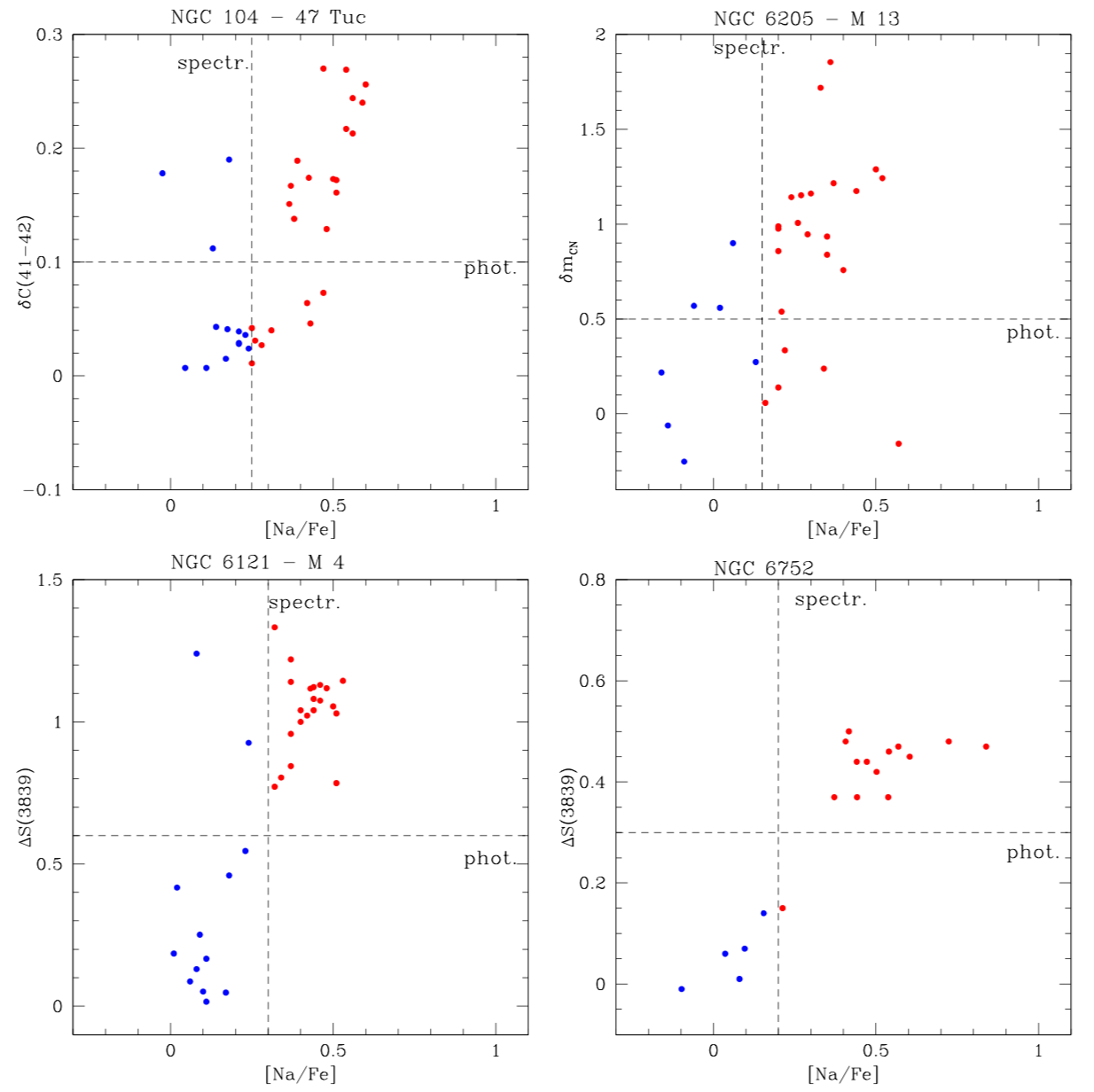}
\caption{The same as in Fig.\ref{f:cpap5} for four more GCs. Upper left panel:
47 Tuc, with CN index from Smith (2015) and O,Na from Cordero et al. (2014).
Upper right panel: M 13, with CN index  from Smith \& Briley (2006) and O,Na
from Johnson \& Pilachowski (2012). Lower left panel: M~4, with CN index from
Smith \&  Briley (2005) and O,Na from Marino et al. (2008). Lower right panel:
NGC~6752,  with CN index from Smith  (2007) and O,Na from Carretta et al.
(2007).}
\label{f:cpap4gc}
\end{figure*}

In Fig.~\ref{f:cpap4gc} the example of NGC~5904 is extended to four other GCs
using a variety of different indicators for CN both from photometric systems or
low resolution spectroscopy in NGC~104, NGC~6205, NGC~6121, and NGC~6752. The
separation between CN-strong and CN-weak stars are from the original papers.
The first three GCs seem to show the same ``pathology" of NGC~5904, with a
discrepant  classification of FG and SG stars in a few cases. In all occurrences
CN-strong stars are found with low levels of Na.

Our definition of second generation(s) is based on the abundance variations of 
Na that distinguish stars with composition altered by proton-capture reactions 
in GCs from their unpolluted counterparts, e.g. in the halo field. The
comparison shown in Fig.~\ref{f:cpap5} and Fig.~\ref{f:cpap4gc} seems to point
out that classification schemes based on CN indices, either photometric or
spectroscopic, tend to overestimate the fraction of FG stars. This result is
well confirmed by the statistics of mismatches in Table~\ref{t:tabmismatches}.
The vast majority of discrepancies ($73\pm 10\%$) consists in stars erroneously 
classified as FG on the PCMs, whereas the contrary is true only for a minor
fraction of the cases ($27\pm 7\%$). 

\begin{figure}
\centering
\includegraphics[scale=0.7]{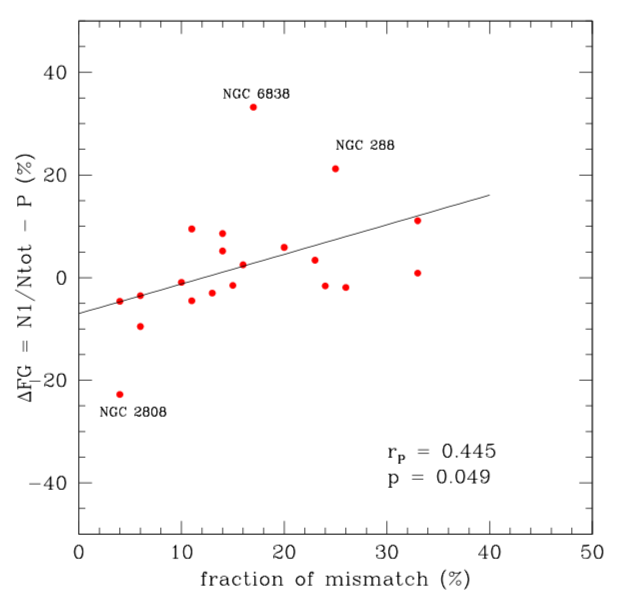}
\caption{Differences $\Delta FG$ in the fraction of FG stars from HST photometry through the
PCMs and from the Na criterion as a function of the fraction of mismatches found
in the present paper. A linear fit, with Pearson correlation coefficient and
probability is also plotted.}
\label{f:diff}
\end{figure}

The impact of this unbalanced discrepancy on the population ratios of MPs in GCs
is visualised in Fig.~\ref{f:diff}, where we plot the difference $\Delta FG$
in the unpolluted FG component as given by HST photometry (N$_1$/N$_{tot}$,
Milone et al. 2017) and as derived from the Na criterion (primordial P 
fraction, Carretta et al. 2010a and references in Tab.~\ref{t:numbers}).

\begin{figure*}
\centering
\includegraphics[scale=0.7]{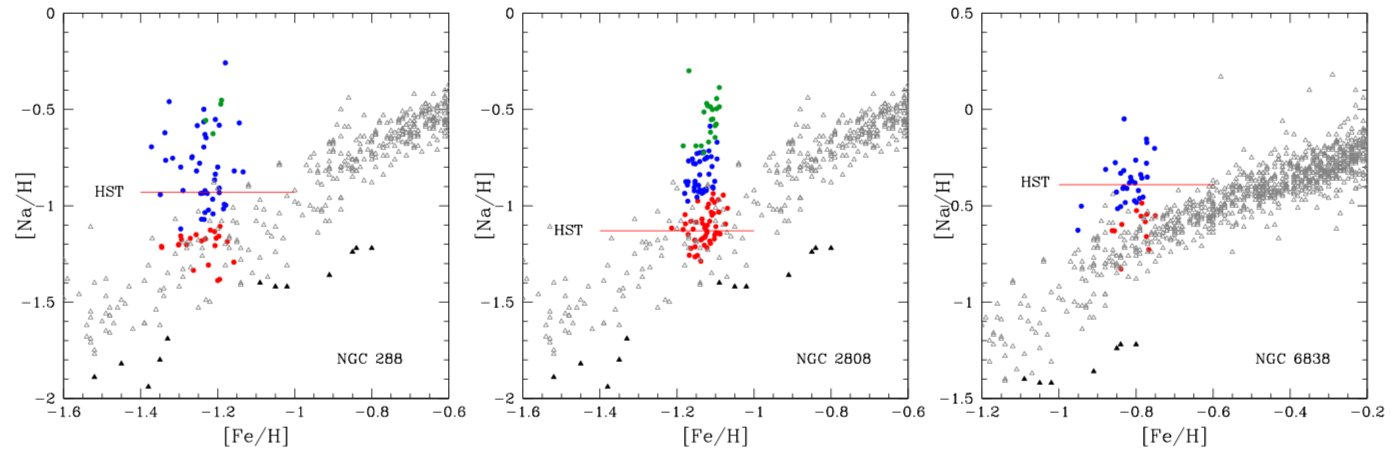}
\caption{Sodium distribution in the three GCs ouliers in Fig.~\ref{f:diff}
compared to the sample of field stars homogenized in Carretta (2013,
triangles). Red, blue, and green filled circles are the P, I, and E components
according to Carretta et al. (2009a). The red line reproduce the fraction of FG
stars according to Milone et al. (2017) from PCMs (see text). From left to right
the panels are for NGC~288, NGC~2808, and NGC~6838.}
\label{f:nah3}
\end{figure*}

\begin{figure}
\centering
\includegraphics[scale=0.8]{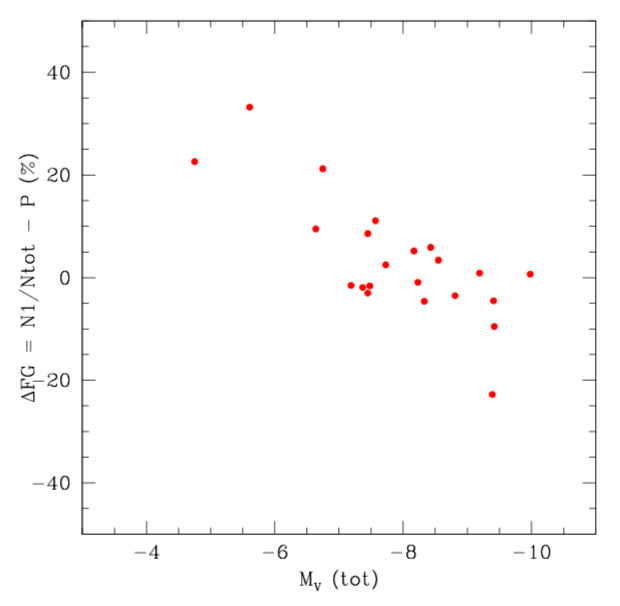}
\caption{Differences $\Delta FG$ as a function of the total absolute magnitude 
of the 22 GCs in this study. }
\label{f:m7b}
\end{figure}

As the fraction of mis-classification increases, the photometric method via PCMs
increasingly overestimates the fraction of FG stars compared to the
spectroscopic approach, and the trend is so regular that a simple linear
regression results statistically significant (the Pearson correlation coefficient
and the two-tail probability are reported in Fig.~\ref{f:diff}).

For the three major outliers in Fig.~\ref{f:diff}, NGC~288, NGC~2808, and
NGC~6838, we compare the spectroscopic and photometric tagging of MPs in
Fig.~\ref{f:nah3}, using the sample of field stars assembled in
Carretta (2013) as reference. This sample is homogenized  to match as much as possible the
abundance analysis of GC stars, so that the FG component in GCs individuated by
Na only must be compatible with the floor of unpolluted field stars. 
In Fig.~\ref{f:nah3}, red, blue, and green filled circles are used for the P, I,
and E components defined in Carretta et al. (2009a). In particular, the
primordial, FG groups in each GC seems to be a fair match for the field
counterpart (empty grey triangles). Filled black triangles indicate stars that 
seem to deviate from the bulk of distribution at a given metallicity. As
discussed in Carretta (2013), they are composed by a sequence of stars whose
origin is attributed to accretion. The horizontal red line in all the panels of
Fig.~\ref{f:nah3} represents the level required by the estimates in Milone et al.
(2017) to reproduce in the spectroscopic sample the fraction of FG stars derived
from the PCMs.

For the cases of NGC~288 (left panel) and NGC~6838 (right panel) the concept of
primordial population according to the N-based selection from HST would include
a consistent fraction of stars whose Na abundances clearly exceed the level
imprinted in field stars by SNe nucleosynthesis only. We remind that several
studies found than only a mere $2-3\%$ of halo stars shows a chemical
composition typical of SG stars in GCs (e.g. Carretta et al. 2010a, Martell et
al. 2011, Koch et al. 2019). On the contrary, in NGC~2808 (middle panel of
Fig.~\ref{f:nah3}) the HST classification would leave out an important number of
FG stars with composition identical to that of unpolluted field stars, in order
to reproduce the low fraction (about $23\%$) quoted in Milone et al. (2017). We
must note that the unusually low FG fraction for NGC~2808 appears to be
at odds with most of the determination from both spectroscopy and photometry
(see Table 9 in Carretta 2015).

How much CN-based evaluations may overestimate the FG fraction with respect to 
the GC global characteristics can be evaluated from Fig.~\ref{f:m7b} where we
show the difference $\Delta FG$ as a function of the total absolute magnitude
for the 22  GCs in the present study.

The total luminosity is a proxy of the present day mass of GCs, but the trend
exists and it is equally statistically significant when using the initial masses
of GCs as listed in the database by H. 
Baumgardt\footnote{\tt \tiny https://people.smp.uq.edu.au/HolgerBaumgardt/globular/parameter.html}.
Our previous studies based on the Na-O anti-correlation detected a
fraction of the primordial P component essentially constant in GCs (about 30\%
of stars), so the trend observed above is hardly surprising, since Milone et al.
(2017) found from their PCM estimates a trend for N$_1$/N$_{tot}$ to increase in
the lower mass GCs. 

To conclude, the N-based methods of population tagging for MPs tend to
overestimate the fraction of FG stars, and this tendency seems to be more 
frequent in more metal-poor and less massive GCs. As a consequence, the fraction
of SG stars in these GCs is underestimated. For the high mass GCs, like e.g.
NGC~2808, this trend could even be reversed, although data are still too
sparse for a firmer conclusion.

The scenario is furthermore made more complex by the need to dilute the matter
processed in proton-capture reactions (in polluters whose exact nature is still
unknown) with pristine gas in GCs (whose abundance can be only
guessed at).

\begin{table*}
\centering
\caption{List of mismatches between the photometric and spectroscopic
classification of MPs}
\begin{tabular}{rccrrcrrl}
\hline
GC & [Na/Fe]$_{\rm min}$ & star ID & $\Delta col$ & $\Delta col3$ & class.& [Na/Fe] & [O/Fe] & mismatch \\
   &   +0.3 dex          &         &              &               &  PCM  &         &        &          \\
\hline
0104 &  0.40	& R0018794 & -0.2206 &  0.2659 &   SG &    0.375 &   0.409 &   yes    \\ 
0288 &  0.15	& R0000592 & -0.2020 &  0.1232 &   FG &    0.198 &   0.040 &   YES    \\ 
0288 &  	& R0000061 & -0.0535 &  0.0978 &   FG &    0.182 &   0.097 &   yes    \\ 
0288 &  	& R0000498 & -0.0508 &  0.0589 &   FG &    0.370 &  -0.023 &   YES    \\ 
0288 &  	& R0001469 & -0.0341 &  0.0491 &   FG &    0.191 &   0.377 &   yes    \\ 
0288 &  	& R0000355 & -0.0268 &  0.0059 &   FG &    0.172 &   0.426 &   yes    \\ 
0288 &  	& R0001265 & -0.0346 & -0.0306 &   FG &    0.247 &   0.302 &   YES    \\ 
0362 & -0.03	& R0005128 & -0.0875 &  0.1806 &   SG &   -0.055 & 	   &   yes    \\ 
0362 &  	& R0000917 & -0.0371 &  0.0196 &   FG &    0.072 &   0.342 &   YES    \\ 
0362 &  	& R0001008 & -0.0290 &  0.0013 &   FG &    0.050 &   0.239 &   yes    \\ 
1851 &  0.00	& R0016222 & -0.0964 &  0.0359 &   FG &    0.133 &   0.282 &   YES    \\ 
2808 &  0.18	& R0009902 & -0.2594 &  0.2921 &   SG &    0.127 &   0.335 &   yes    \\ 
2808 &  	& R0033840 & -0.2369 &  0.2102 &   SG &    0.170 & 	   &   yes    \\ 
3201 &  0.00	& R0002194 & -0.2427 &  0.0767 &   FG &    0.057 &  -0.086 &   YES    \\ 
3201 &  	& R0003189 & -0.0151 & -0.0062 &   FG &    0.135 & 	   &   YES    \\ 
3201 &  	& R0003013 &  0.0098 & -0.0187 &   FG &    0.016 &   0.191 &   YES    \\ 
3201 &  	& R0000409 &  0.0670 & -0.0208 &   FG &    0.001 &   0.373 &   yes    \\ 
4590 &  0.25	& R0000498 & -0.0535 &  0.0288 &   FG &    0.260 & 	   &   yes    \\ 
4590 &  	& R0000791 & -0.0441 &  0.0156 &   FG &    0.368 &   0.299 &   YES    \\ 
4590 &  	& R0000383 & -0.0389 &  0.0579 &   SG &    0.080 & 	   &   YES    \\ 
4590 &  	& R0000899 & -0.0371 &  0.0826 &   SG &    0.111 & 	   &   YES    \\ 
4590 &  	& R0000317 & -0.0613 &  0.0917 &   SG &    0.199 &   0.040 &   YES    \\ 
4590 &  	& R0001167 & -0.0489 &  0.1196 &   SG &    0.193 & 	   &   yes    \\ 
4833 &  0.29	& R0003717 & -0.0282 &  0.0634 &   FG &    0.597 &   0.382 &   YES    \\ 
4833 &  	& R0001553 & -0.0399 &  0.0452 &   FG &    0.293 &   0.273 &   YES    \\ 
5904 &  0.10	& R0012503 & -0.0041 &  0.1754 &   SG &    0.098 &   0.494 &   yes    \\ 
6093 &  0.25	& R0002835 & -0.0953 &  0.0413 &   FG &    0.393 &   0.200 &   YES    \\ 
6093 &  	& R0000925 & -0.0187 &  0.0858 &   FG &    0.695 &   0.415 &   YES    \\ 
6093 &  	& R0004803 & -0.0803 &  0.0944 &   FG &    0.478 &   0.127 &   YES    \\ 
6121 &  0.25	& R0000553 & -0.0483 &  0.1091 &   SG &    0.20  &   0.48  &   YES    \\ 
6121 &  	& R0001582 & -0.0461 &  0.1149 &   SG &    0.11  &   0.51  &   YES    \\ 
6121 &  	& R0001355 & -0.0775 &  0.1195 &   SG &    0.23  &   0.40  &   YES    \\ 
6121 &  	& R0001022 & -0.0755 &  0.1276 &   SG &    0.25  &   0.38  &   YES    \\ 
6205 &  0.00	& R0003306 &  0.0104 &  0.1905 &   SG &   -0.14  &   0.09  &   YES    \\ 
6205 &  	& R0014192 & -0.0910 &  0.0283 &   FG &    0.45  &   0.12  &   YES    \\ 
6205 &  	& R0015663 & -0.0087 &  0.0272 &   FG &    0.08  &   0.25  &   YES    \\ 
6205 &  	& R0011115 & -0.0106 &  0.0434 &   FG &    0.04  &   0.36  &   yes    \\ 
6205 &  	& R0003505 &  0.0015 &  0.0460 &   FG &    0.22  &   0.30  &   YES    \\ 
6205 &  	& R0014143 & -0.0647 &  0.0207 &   FG &    0.09  &   0.34  &   YES    \\ 
6254 &  0.02	& R0004141 & -0.2074 &  0.0550 &   FG &    0.182 &   0.309 &   YES    \\ 
6254 &  	& R0002170 & -0.1166 &  0.0188 &   FG &    0.073 & 	   &   yes    \\ 
6254 &  	& R0001688 & -0.0438 & -0.0130 &   FG &    0.097 & 	   &   YES    \\ 
6254 &  	& R0004487 &  0.0082 &  0.0372 &   FG &    0.178 &   0.473 &   YES    \\ 
6254 &  	& R0003806 & -0.0795 &  0.0623 &   FG &    0.076 &   0.304 &   YES    \\ 
6254 &  	& R0001328 & -0.0419 &  0.0223 &   FG &    0.181 &   0.529 &   YES    \\ 
6388 & 0.20     & R0018623 &  0.0059 &  0.7061 &   FG &    0.111 &  -0.149 &   YES    \\ 
6388 &          & R0023149 & -0.4395 &  0.3905 &   FG &    0.093 &   0.089 &   YES    \\ 
6388 &          & R0011104 & -0.2750 &  0.3736 &   FG &    0.134 &   0.353 &   YES    \\ 
6388 &          & R0022136 & -0.2303 &  0.1892 &   FG &   -0.014 &   0.010 &   YES    \\ 
6397 &  0.10	& R0001425 & -0.0466 &  0.0587 &   SG &   -0.134 & 	   &   YES    \\ 
6397 &  	& R0000935 & -0.0220 &  0.0228 &   FG &    0.118 & 	   &   yes    \\ 
6752 &  0.20	& R0003195 & -0.0350 &  0.1662 &   SG &    0.120 &   0.695 &   YES    \\ 
6752 &  	& R0004277 & -0.0182 & -0.0065 &   FG &    0.365 &   0.523 &   YES    \\ 
6752 &  	& R0007736 &  0.0217 &  0.0129 &   FG &    0.648 & 	   &   YES    \\ 
6752 &  	& R0005094 & -0.0363 &  0.0540 &   FG &    0.338 &   0.183 &   YES    \\ 
6809 &  0.05	& R0001112 & -0.0518 &  0.0655 &   FG &    0.613 &   0.212 &   YES    \\ 
6809 &  	& R0001928 & -0.0210 &  0.0604 &   FG &    0.875 &   0.082 &   YES    \\ 
6809 &  	& R0000880 & -0.0169 &  0.0617 &   FG &    0.413 &   0.081 &   YES    \\ 
6809 &  	& R0000087 & -0.0041 &  0.0620 &   FG &    0.539 &   0.280 &   YES    \\ 
6809 &  	& R0001035 & -0.0171 &  0.0444 &   FG &    0.364 &   0.235 &   YES    \\ 
6809 &  	& R0002430 & -0.0051 &  0.0251 &   FG &    0.210 &   0.410 &   YES    \\ 
6809 &  	& R0002134 & -0.0012 &  0.0027 &   FG &    0.260 &   0.429 &   YES    \\ 
\hline 
\end{tabular}
\end{table*}

\addtocounter{table}{-1}

\begin{table*}[t!]
\centering
\caption{continue}
\begin{tabular}{rccrrcrrl}
\hline

GC & [Na/Fe]$_{\rm min}$ & star ID & $\Delta col$ & $\Delta col3$ & class.& [Na/Fe] & [O/Fe] & mismatch \\
   &   +0.3 dex          &         &              &               &  PCM  &         &        &          \\
\hline
6838 &  0.30	& R0001028 & -0.1492 &  0.2445 &   SG &    0.226 &   0.325 &   yes    \\  
6838 &  	& R0000100 & -0.0825 &  0.0189 &   FG &    0.373 &   0.418 &   yes    \\  
6838 &  	& R0000419 & -0.0600 &  0.0171 &   FG &    0.316 &   0.350 &   yes    \\  
6838 &  	& R0001862 & -0.0125 &  0.0158 &   FG &    0.436 &   0.424 &   yes    \\  
7078 &  0.22	& R0000861 & -0.1049 &  0.1551 &   SG &    0.166 & 	   &   yes    \\  
7078 &  	& R0000106 & -0.1521 &  0.1501 &   SG &    0.188 & 	   &   yes    \\  
7078 &  	& R0008536 & -0.2064 &  0.0433 &   FG &    0.440 &   0.242 &   YES    \\  
7078 &  	& R0001894 & -0.0725 &  0.0425 &   FG &    0.643 & 	   &   YES    \\  
7078 &  	& R0007957 & -0.0646 &  0.0328 &   FG &    0.293 &   0.262 &   YES    \\  
7078 &  	& R0002778 & -0.0900 &  0.0217 &   FG &    0.375 & 	   &   YES    \\  
7078 &  	& R0006154 & -0.0065 &  0.0057 &   FG &    0.266 & 	   &   yes    \\  
7099 &  0.22	& R0002313 & -0.0831 &  0.1309 &   SG &    0.099 &   0.506 &   YES    \\  
7099 &  	& R0003697 & -0.0196 &  0.0007 &   FG &    0.461 &   0.710 &   YES    \\  

\hline
\end{tabular}
\label{t:tabmismatches}
\end{table*}

\begin{table}[h]
\centering
\caption{Number of mismatches between the photometric and spectroscopic
classification of MPs}
\begin{tabular}{ll|lc}
\hline
GC  &  mismatch &  GC   & mismatch \\
\hline
NGC~~104 & 1 (6\%)  & NGC~6205 & 6 (23\%) \\  
NGC~~288 & 6 (25\%) & NGC~6254 & 6 (24\%) \\  
NGC~~362 & 3 (20\%) & NGC~6388 & 4 (11\%) \\  
NGC~1851 & 1 (4\%)  & NGC~6397 & 2 (11\%) \\  
NGC~2808 & 2 (4\%)  & NGC~6535 & 0  \\        
NGC~3201 & 4 (14\%) & NGC~6715 & 0  \\        
NGC~4590 & 6 (26\%) & NGC~6752 & 4 (16\%) \\  
NGC~4833 & 2 (14\%) & NGC~6809 & 7 (33\%) \\  
NGC~5904 & 1 (6\%)  & NGC~6838 & 4 (17\%) \\  
NGC~6093 & 3 (10\%) & NGC~7078 & 7 (33\%) \\  
NGC~6121 & 4 (15\%) & NGC~7099 & 2 (13\%) \\  
\hline
\end{tabular}
\label{t:totmismatch}
\end{table}

\section{Summary}

We addressed the issue of population tagging of MPs in GCs through different
methods. To this purpose we compared indicators based on HST photometry and
direct probes of MPs such as the Na abundance from high resolution spectroscopy.

Due to the lack of published data, we newly derived pseudo-colour maps (PCMs) for
22 GCs with photometric  catalogues from Nardiello et al. (2018). To check how
photometric indicators perform in the population tagging, the photometry was
then matched with stars having Na abundances mostly from our very homogeneous
FLAMES survey. With respect to a similar attempt by Marino et al. (2019) we
retrieved all the stars they used with both HST photometry and high resolution
spectroscopy available and added a new sample of matches, doubling the database
of cross identifications between photometry and spectroscopy.

Our conclusions can be summarised as follows:
\begin{itemize}
\item[1)] Our analysis shows that PCMs perform a coarsely correct ranking in
populations for the bulk of stars in GCs, based essentially on the pseudo-colour
$col3$ sensitive basically to N abundances through the effect of NH and CN
molecular bands on fluxes in proper bandpasses.
\item[2)] When the classification from PCM is compared to the direct
individuation of MPs from spectroscopic abundances, there are a number of
mismatches or mis-classifications, in particular when Na is used as privileged
species to disentangle FG and SG stars. Stars populating the SG region are found
with low Na abundances whereas stars belonging to the PCM region of FG stars
have Na abundances at the level of stars with content of light elements clearly
altered by proton-capture reactions responsible for the MPs phenomenon. The
latter group constitutes the majority of the observed mis-classifications.
\item[3)] On average, we found that about $16\%$
of stars with spectroscopic abundances are erroneously classified in PCMs. The
fraction of mismatches increases in metal-poor GCs, suggesting that the
weakening of molecular features may play a role in the
mis-classification.
\item[4)] Hence, the estimates of the fraction of FG stars are different when
using different indicators. Those based on NH or CN (hence the majority from
UV/optical photometry) are usually higher than those from pure spectroscopic
abundances from heavier elements. The comparison with the unpolluted floor as
given by halo field stars shows that photometric tagging tends to overestimate
the FG fraction.
\item[5)] The difference in estimates seems to vary from cluster to cluster,
according to the cluster total mass. The overestimate of the FG fraction seems
to be quantitatively more severe and pronounced in more metal-poor and less
massive GCs.

\end{itemize}

To conclude, what we found in the present work is the modern form assumed by 
the well known decoupling observed in abundances of N and Na among the MPs in
GCs. We confirm the early extensive studies by Smith and collaborators: the
phenomenon as seen by CN, and O, Na is the same, namely the multiple stellar
populations in GCs, but different aspects are likely sampled by the lighter and
the heavier proton-capture elements. 

The PCMs seem to follow rather well only the outcome of the C-N cycle at 
moderate or low temperatures ($T > 10$ MK for C$\rightarrow$N processing). 
When going to  higher temperature regimes, where both the O-N cycle and the 
Ne-Na chain are at work ($T > 40$ MK), there are mistakes in the classification
and usually stars with Na abundances high enough to be located in the SG region
along the Na-O anti-correlation are mistaken for FG stars.

This finding represents a strong caveat for all the issues concerning the
correct population ratios in MPs, like e.g. the mass budget problem, since the 
FG fraction is the pool from which some massive polluters provide the nuclearly
processed matter to be used, together with pristine gas, for the assembly of the
stellar population with altered chemical composition. Despite some decades of
observations from both photometry and spectroscopy gained for GC stars, the
whole issue of MPs seems to be even now still far to be solved. 

Finally, we note that all the conclusions in the present work have been possible
only thanks to publicly available data.

\begin{acknowledgements}
This research has made large use of the SIMBAD database (in
particular  Vizier), operated at CDS, Strasbourg, France, of the NASA's
Astrophysical Data System, and TOPCAT.
\end{acknowledgements}

\FloatBarrier

\begin{appendix}
\onecolumn
\section{Pseudo-colour maps of all the GCs in the sample}

In this Appendix we present in detail the results of the procedure illustrated
in Section 2 for the remaining of the GCs in the sample not shown in
Fig.~\ref{f:4pannelli}. 

\begin{figure*}
\centering
\includegraphics[scale=1.1]{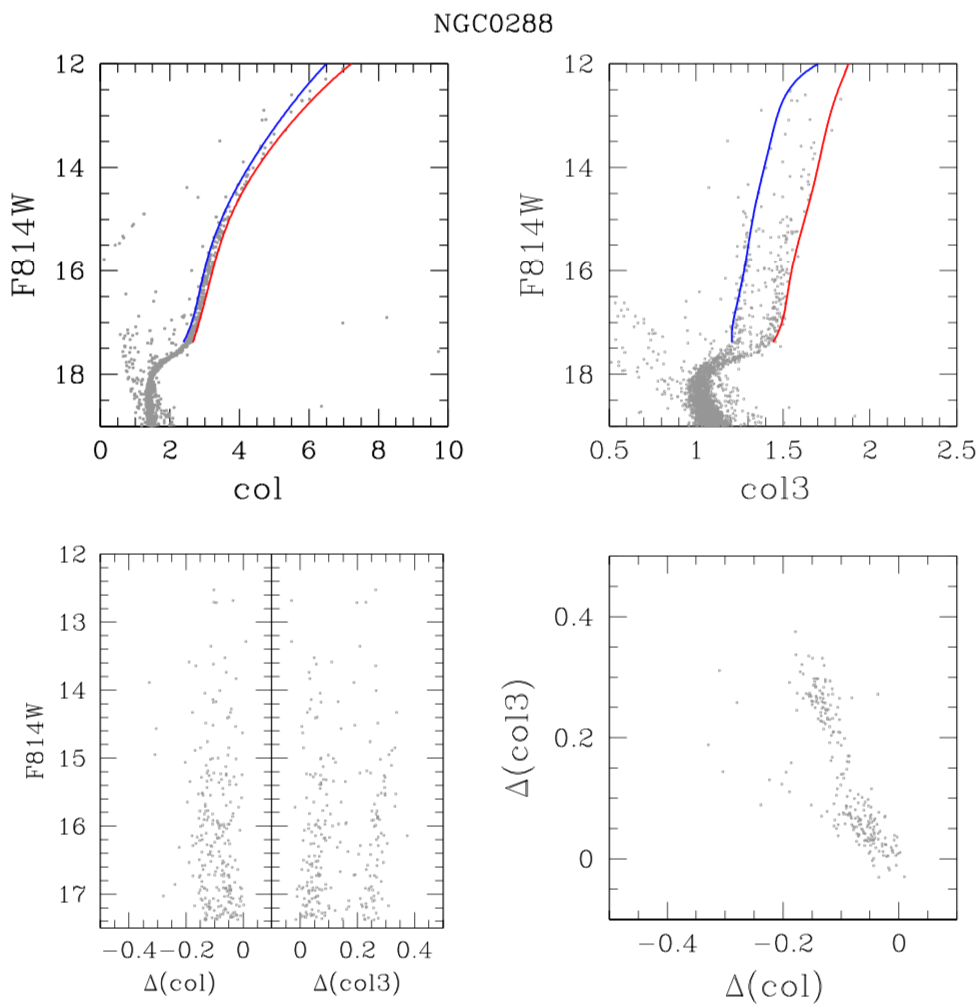}
\caption{As in Fig. 1 but for the globular cluster NGC~288}
\label{f:app0288}
\end{figure*}

\begin{figure*}
\centering
\includegraphics[scale=1.1]{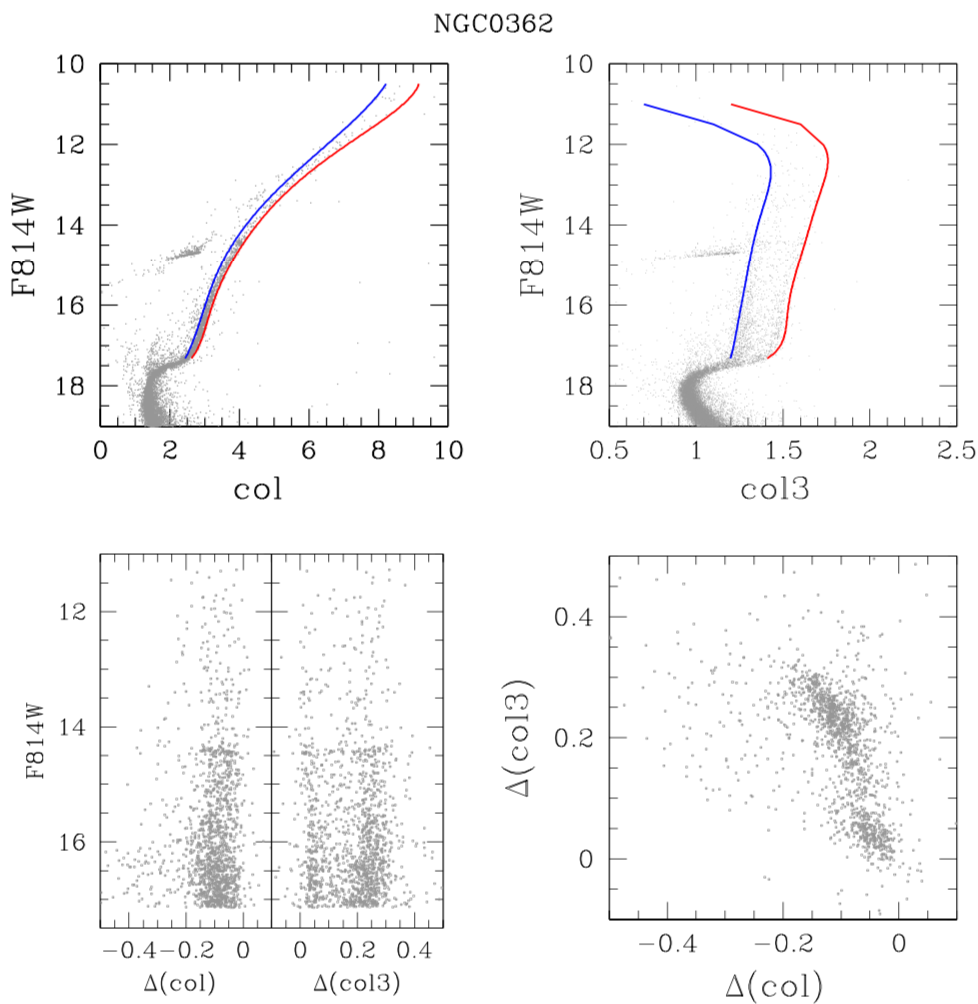}
\caption{As in Fig. 1 but for the globular cluster NGC~362}
\label{f:app0362}
\end{figure*}

\begin{figure*}
\centering
\includegraphics[scale=1.1]{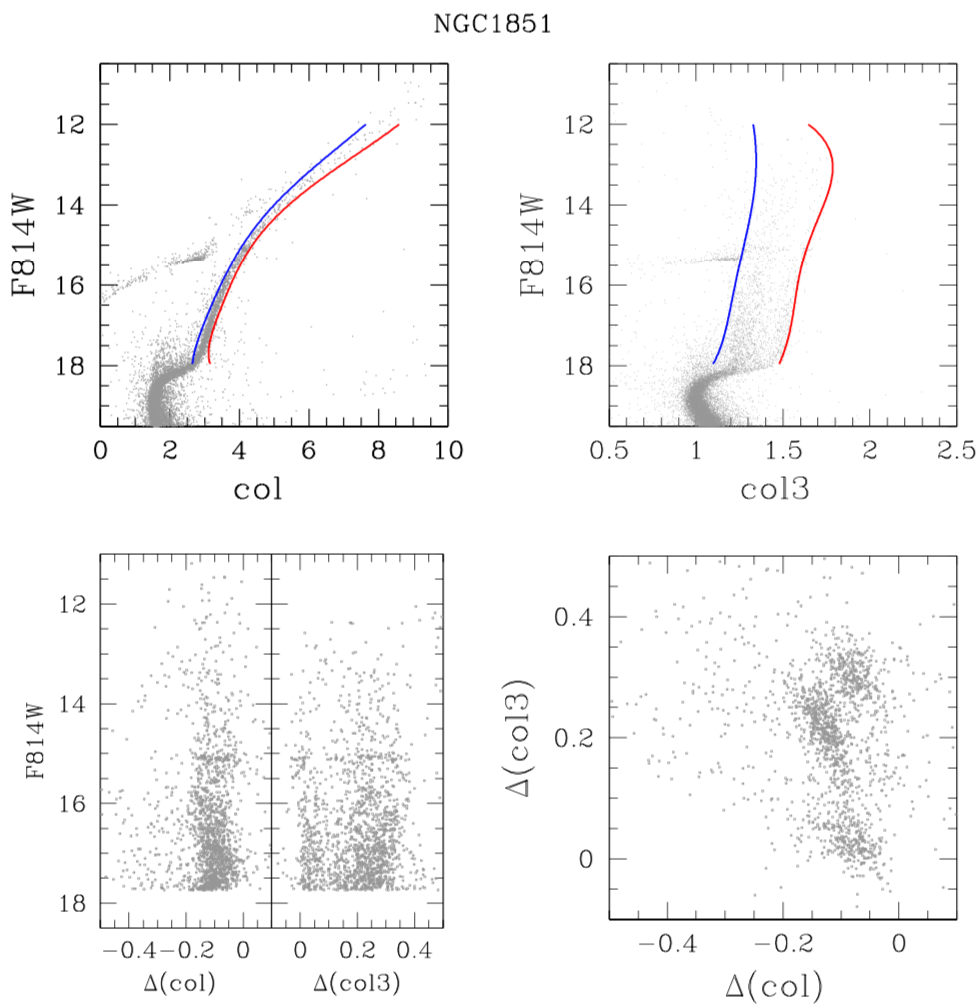}
\caption{As in Fig. 1 but for the globular cluster NGC~1851}
\label{f:app1851}
\end{figure*}

\begin{figure*}
\centering
\includegraphics[scale=1.1]{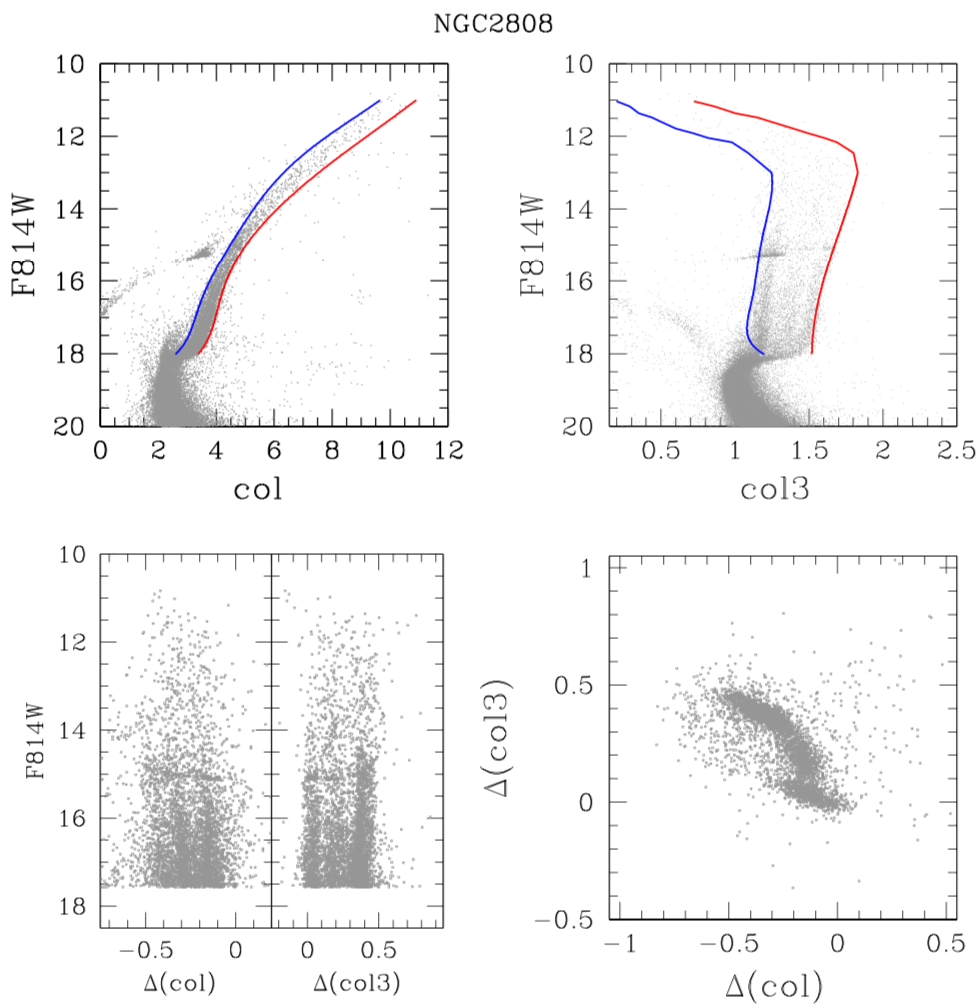}
\caption{As in Fig. 1 but for the globular cluster NGC~2808}
\label{f:app2808}
\end{figure*}

\begin{figure*}
\centering
\includegraphics[scale=1.1]{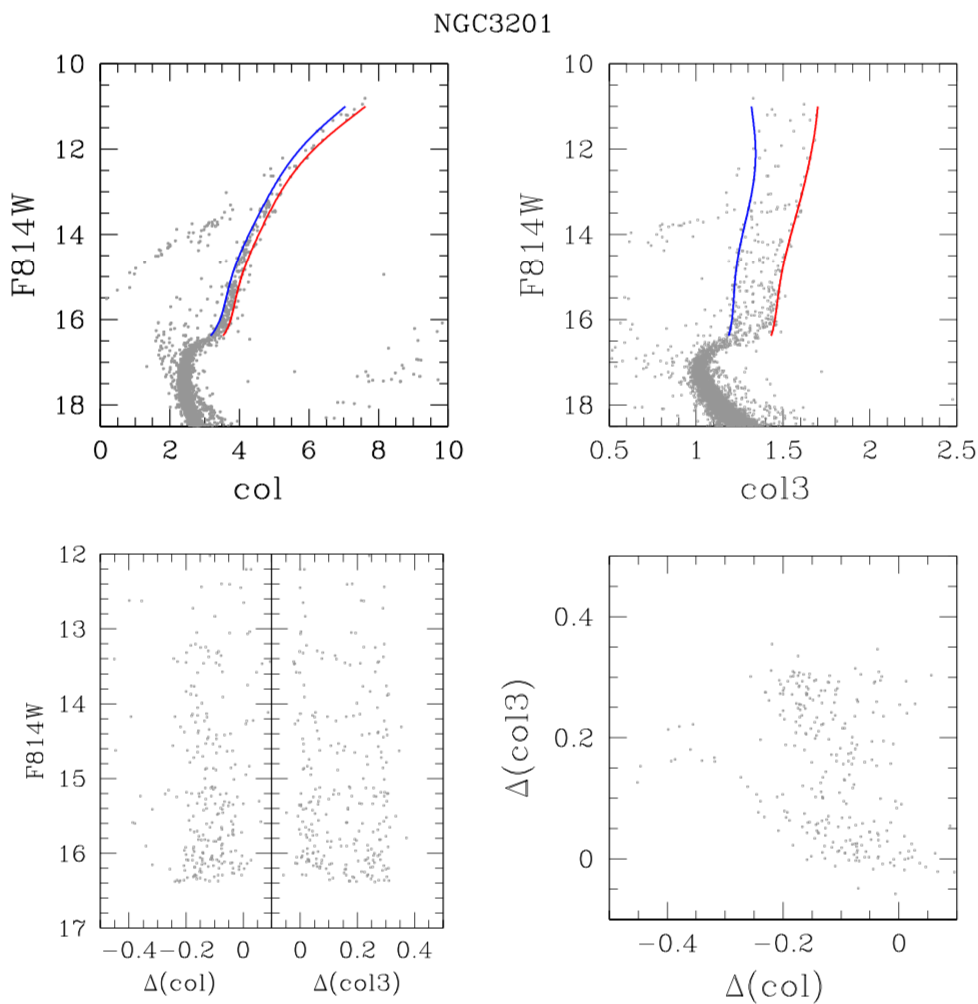}
\caption{As in Fig. 1 but for the globular cluster NGC~3201}
\label{f:app3201}
\end{figure*}

\begin{figure*}
\centering
\includegraphics[scale=1.1]{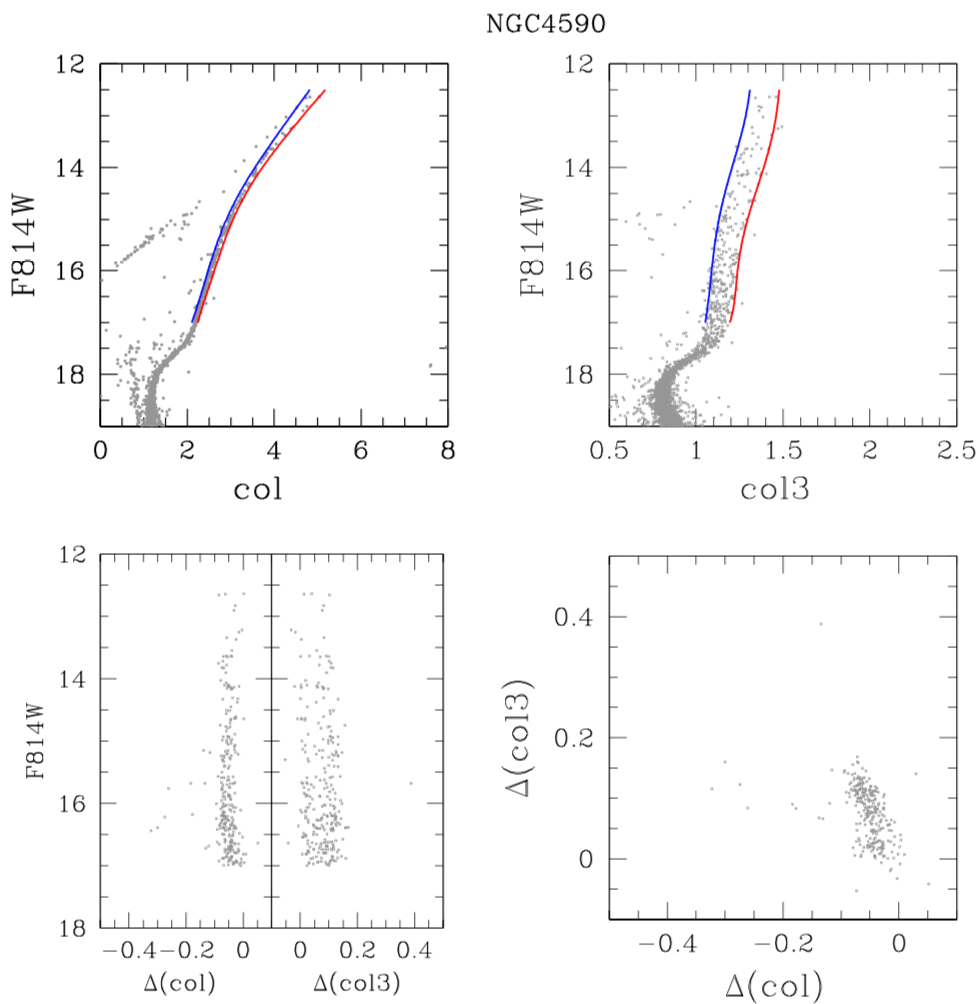}
\caption{As in Fig. 1 but for the globular cluster NGC~4590}
\label{f:app4590}
\end{figure*}

\begin{figure*}
\centering
\includegraphics[scale=1.1]{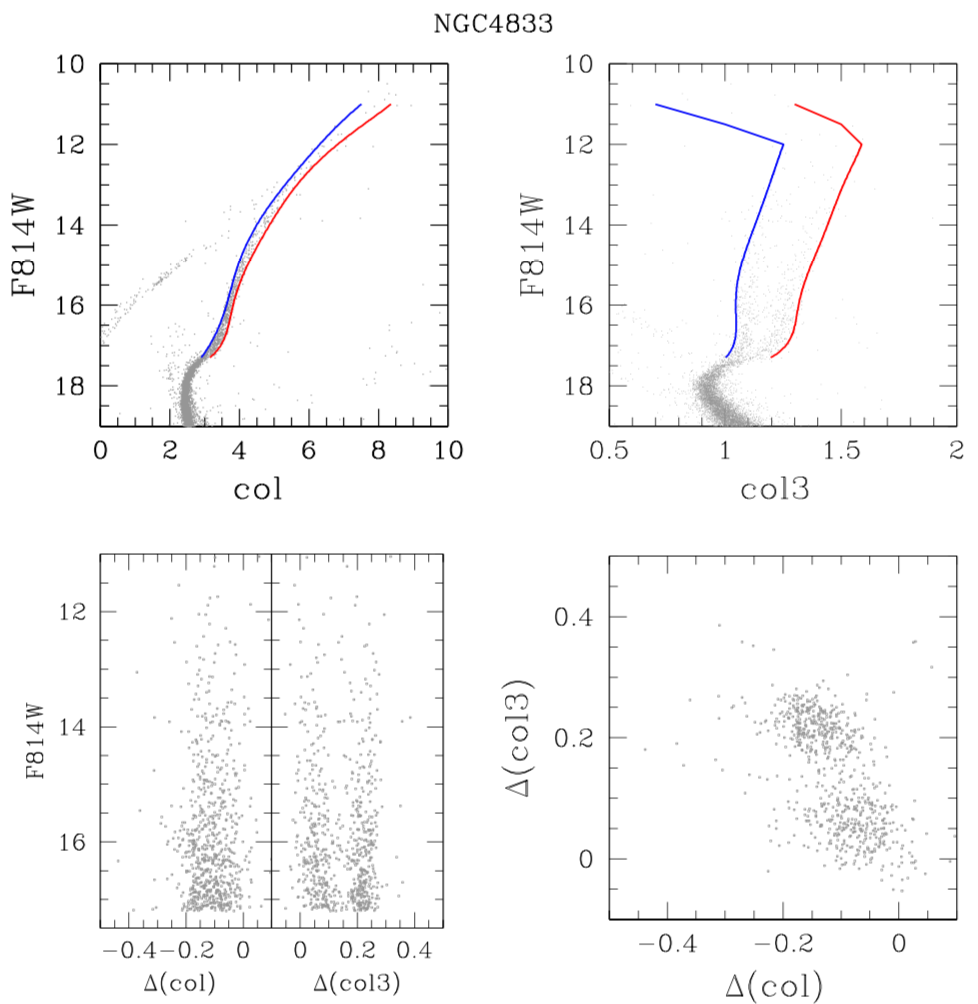}
\caption{As in Fig. 1 but for the globular cluster NGC~4833}
\label{f:app4833}
\end{figure*}

\begin{figure*}
\centering
\includegraphics[scale=1.1]{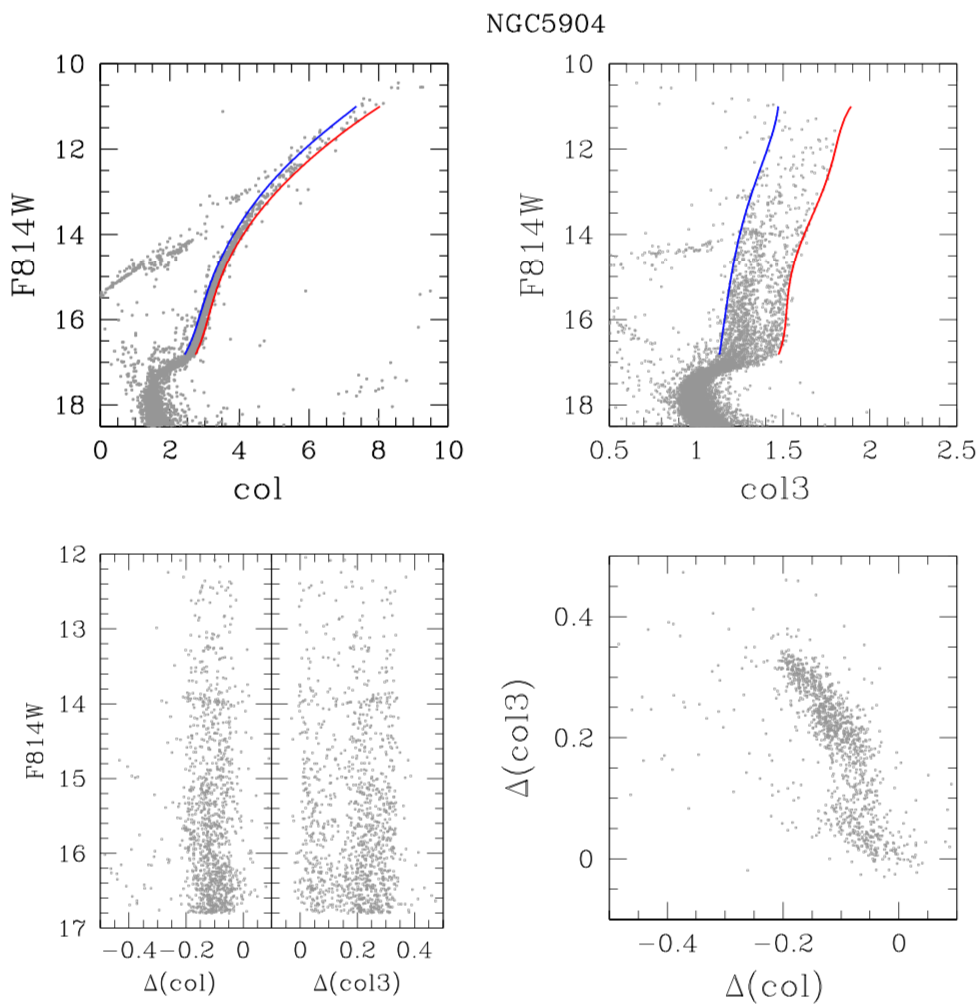}
\caption{As in Fig. 1 but for the globular cluster NGC~5904}
\label{f:app5904}
\end{figure*}

\begin{figure*}
\centering
\includegraphics[scale=1.1]{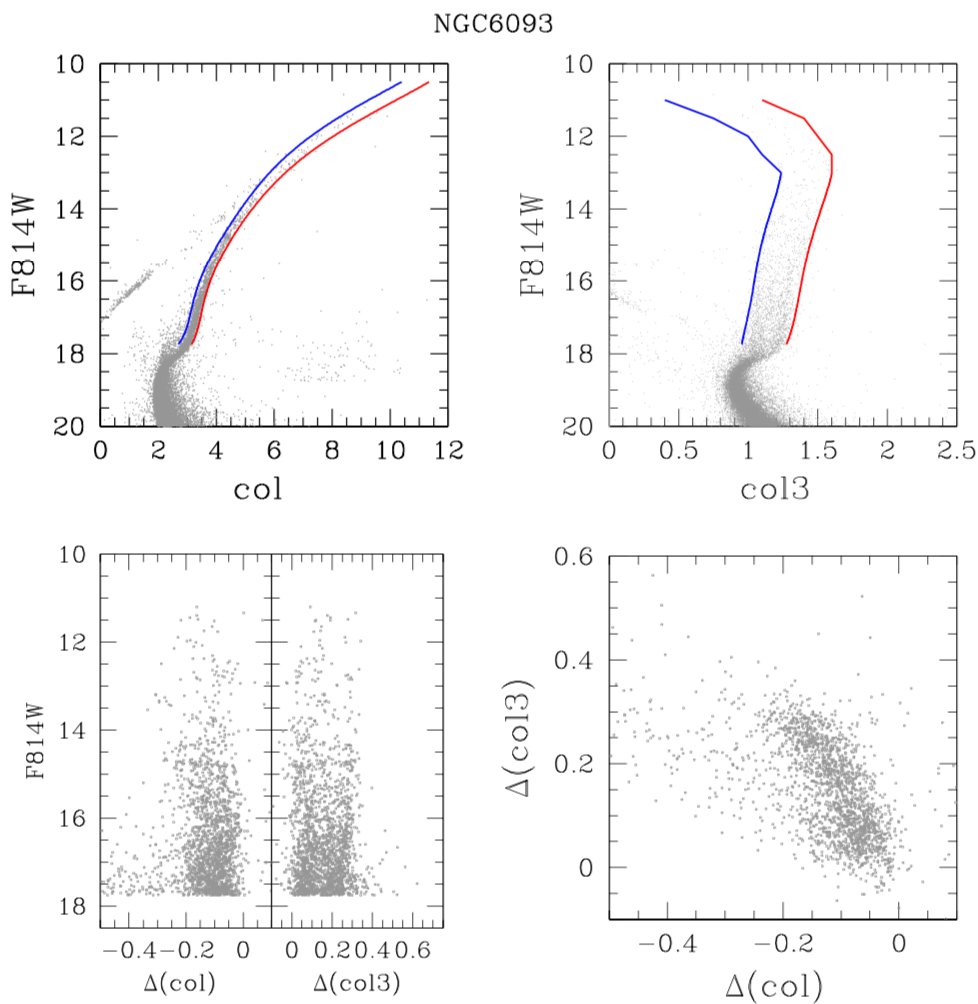}
\caption{As in Fig. 1 but for the globular cluster NGC~6093}
\label{f:app6093}
\end{figure*}

\begin{figure*}
\centering
\includegraphics[scale=1.1]{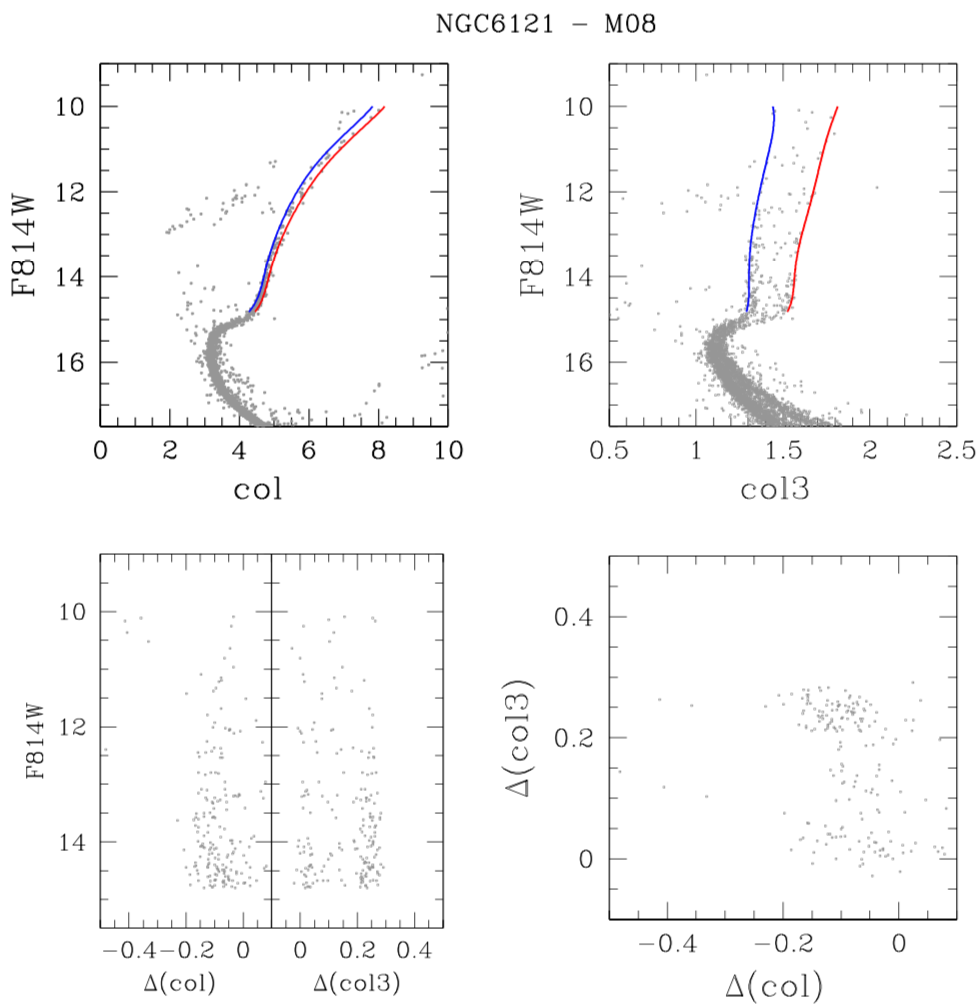}
\caption{As in Fig. 1 but for the globular cluster NGC~6121}
\label{f:app6121}
\end{figure*}

\begin{figure*}
\centering
\includegraphics[scale=1.1]{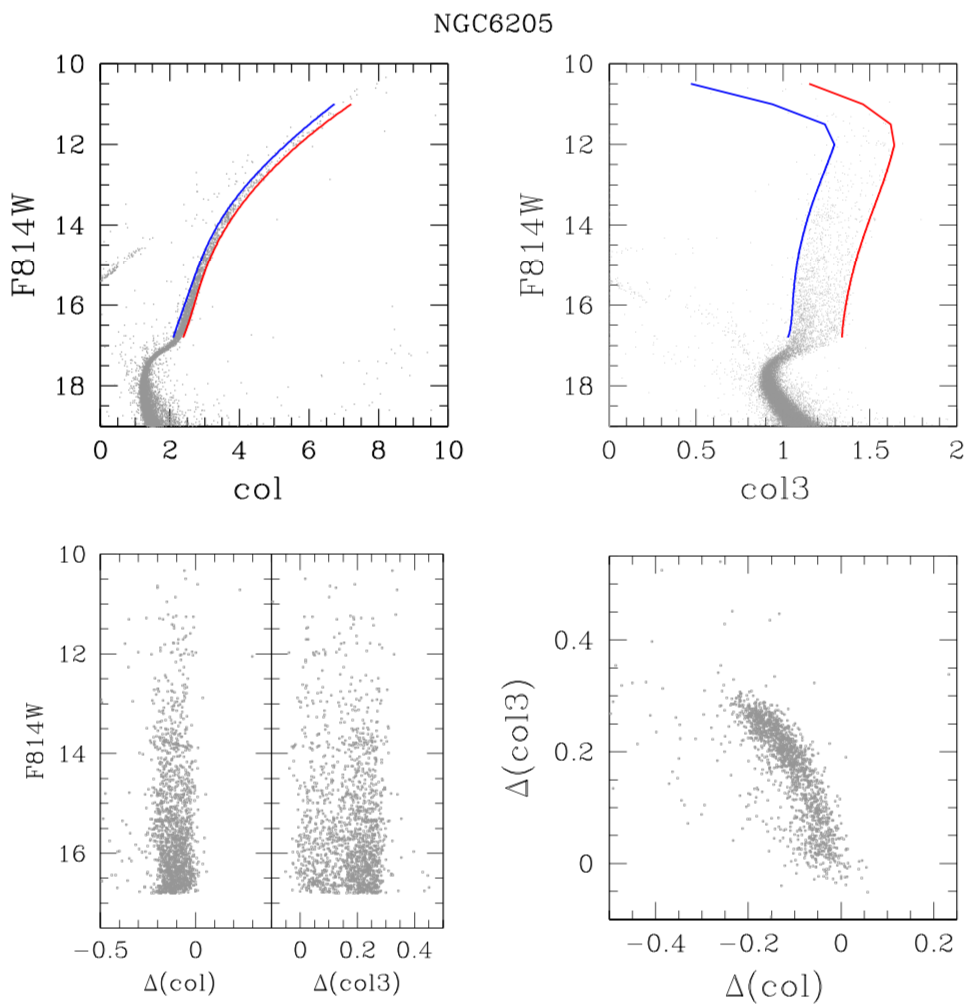}
\caption{As in Fig. 1 but for the globular cluster NGC~6205}
\label{f:app6205}
\end{figure*}

\begin{figure*}
\centering
\includegraphics[scale=1.1]{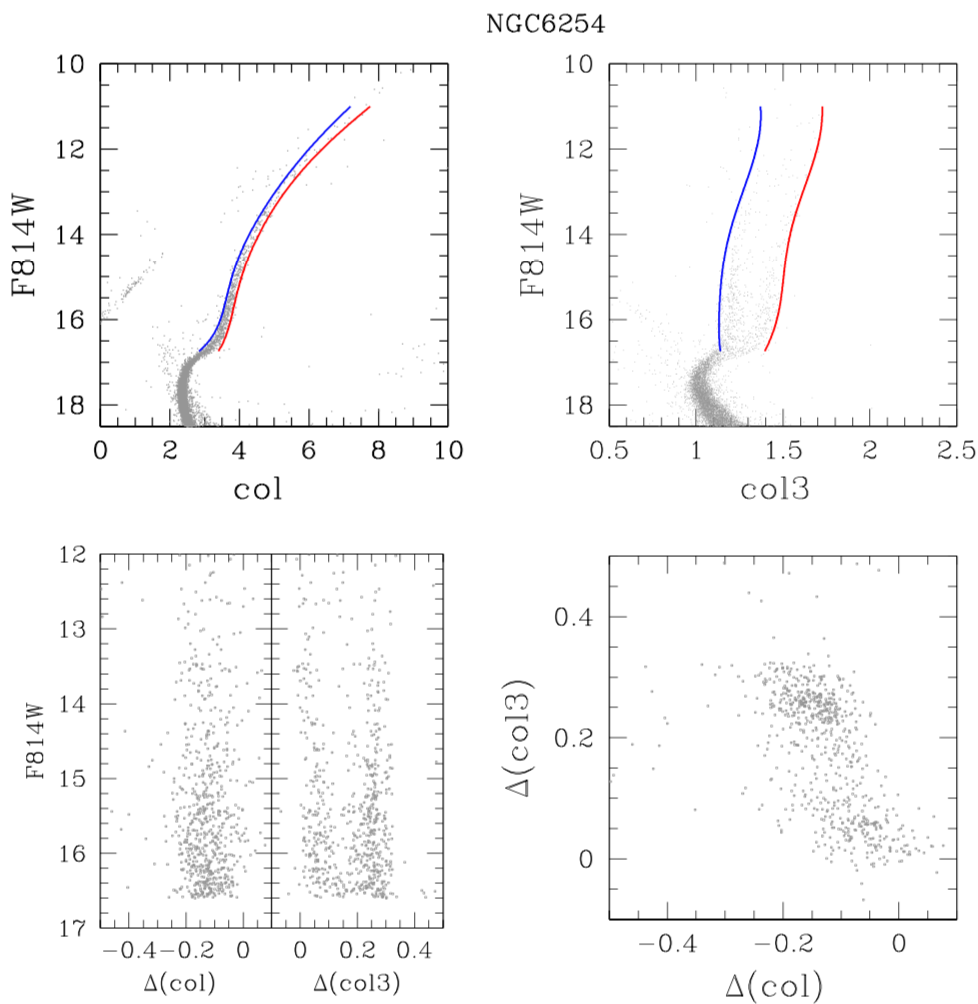}
\caption{As in Fig. 1 but for the globular cluster NGC~6254}
\label{f:app6254}
\end{figure*}

\begin{figure*}
\centering
\includegraphics[scale=1.1]{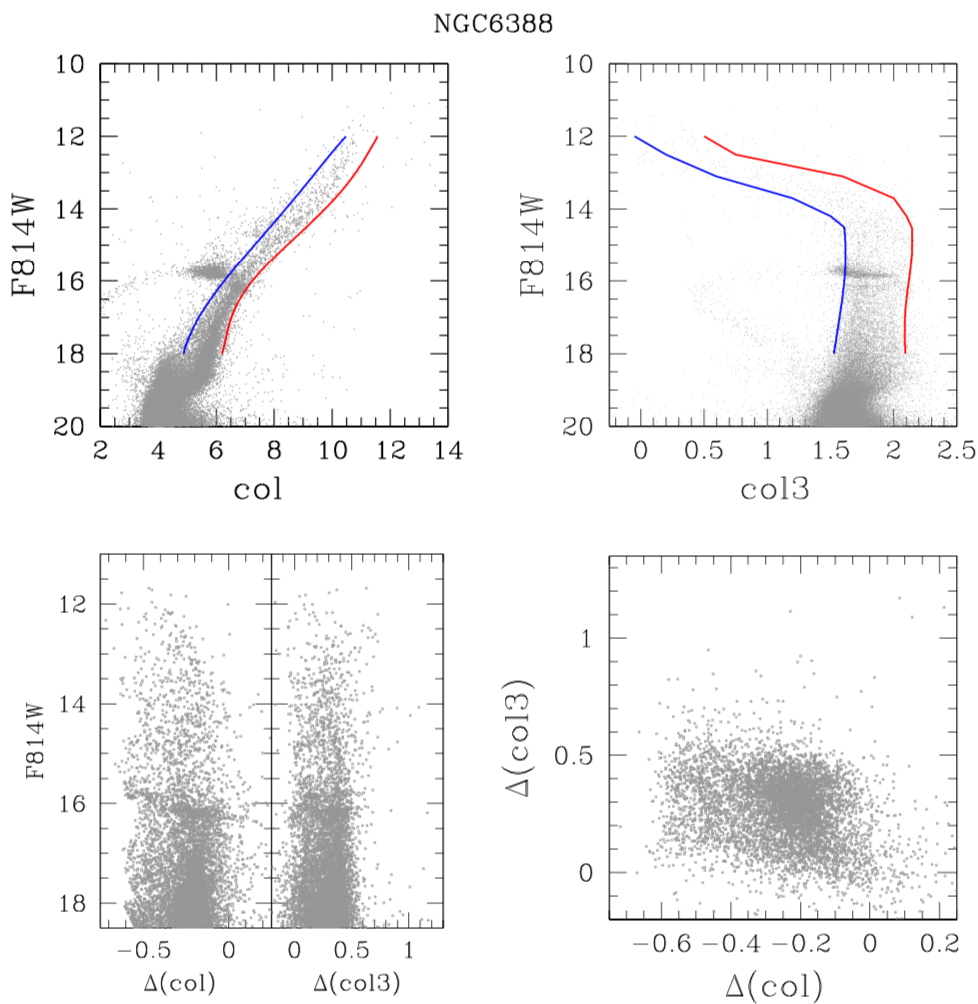}
\caption{As in Fig. 1 but for the globular cluster NGC~6388}
\label{f:app6388}
\end{figure*}

\begin{figure*}
\centering
\includegraphics[scale=1.1]{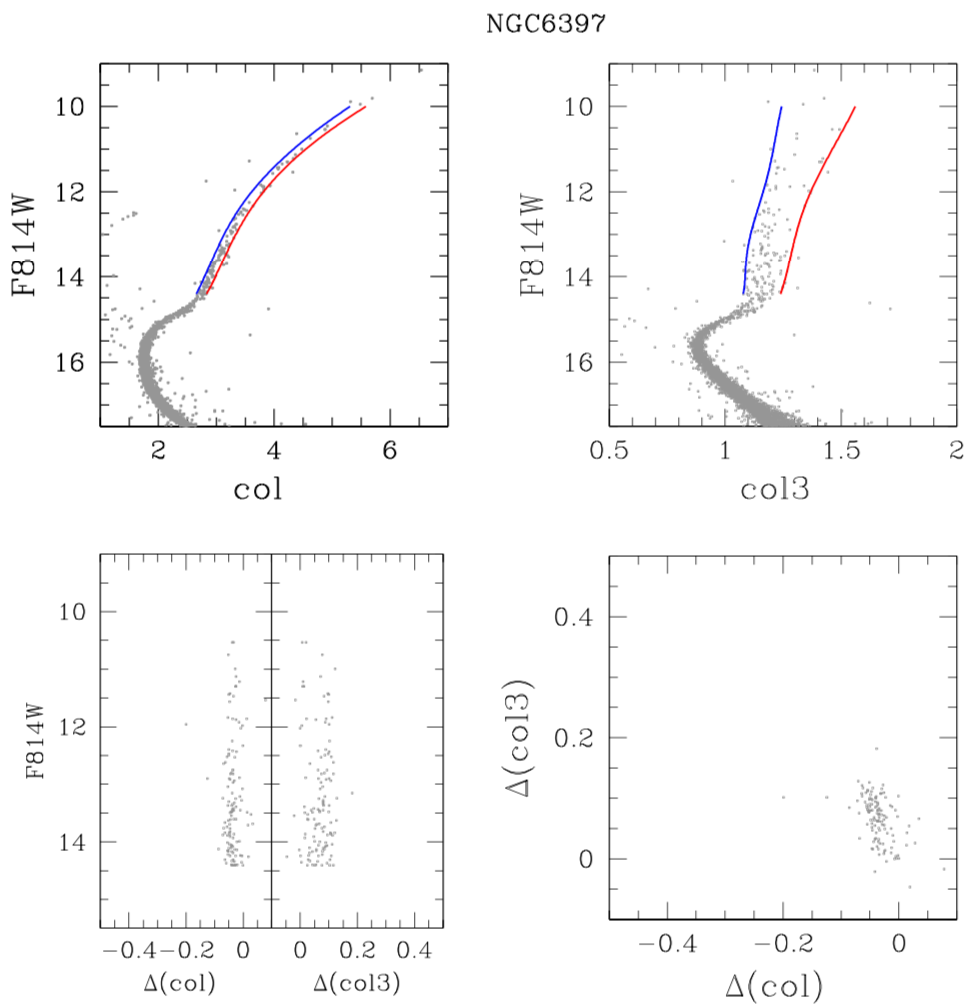}
\caption{As in Fig. 1 but for the globular cluster NGC~6397}
\label{f:app6397}
\end{figure*}

\begin{figure*}
\centering
\includegraphics[scale=1.1]{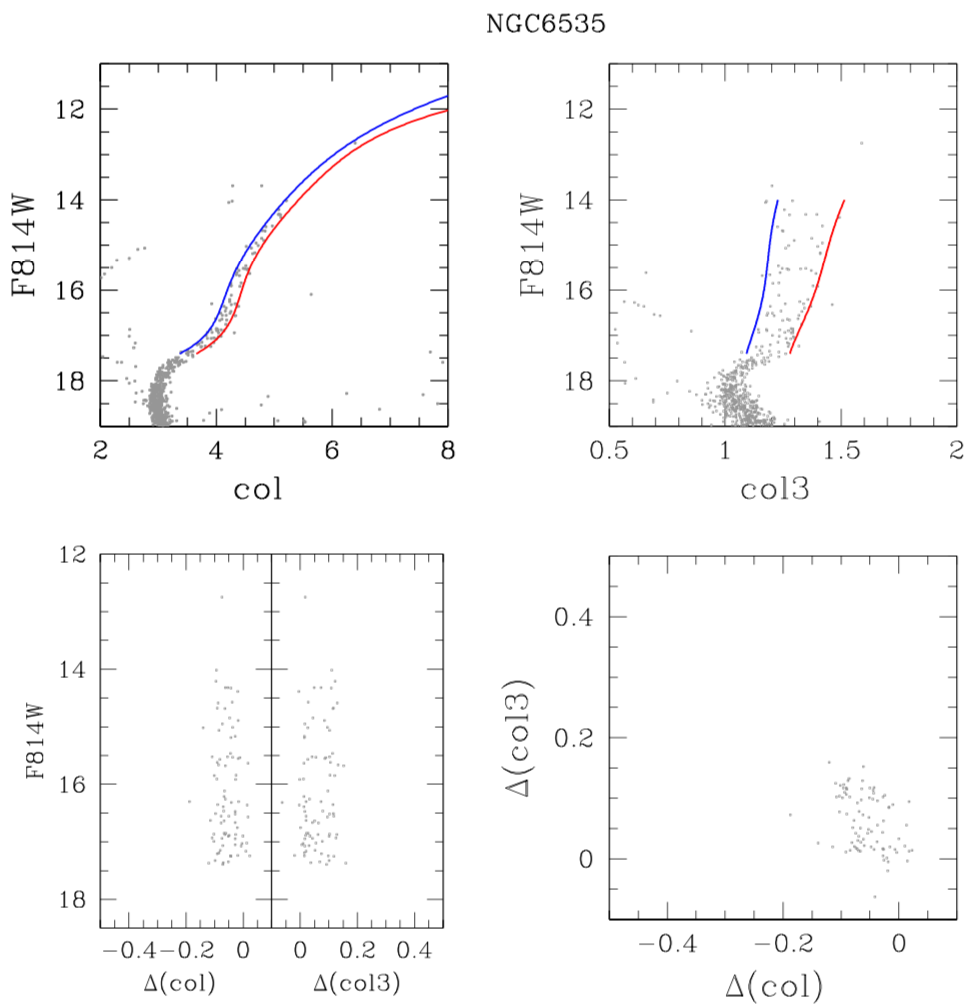}
\caption{As in Fig. 1 but for the globular cluster NGC~6535}
\label{f:app6535}
\end{figure*}

\begin{figure*}
\centering
\includegraphics[scale=1.1]{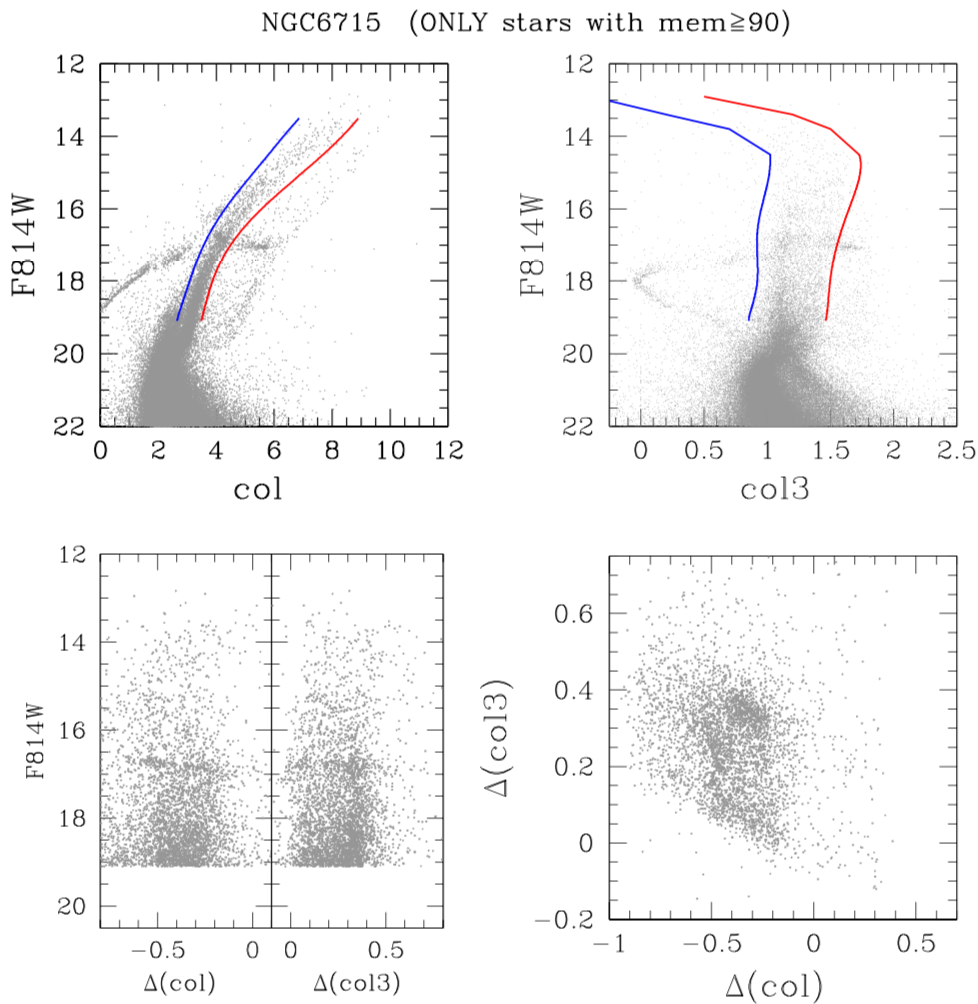}
\caption{As in Fig. 1 but for the globular cluster NGC~6715}
\label{f:app6715}
\end{figure*}

\begin{figure*}
\centering
\includegraphics[scale=1.1]{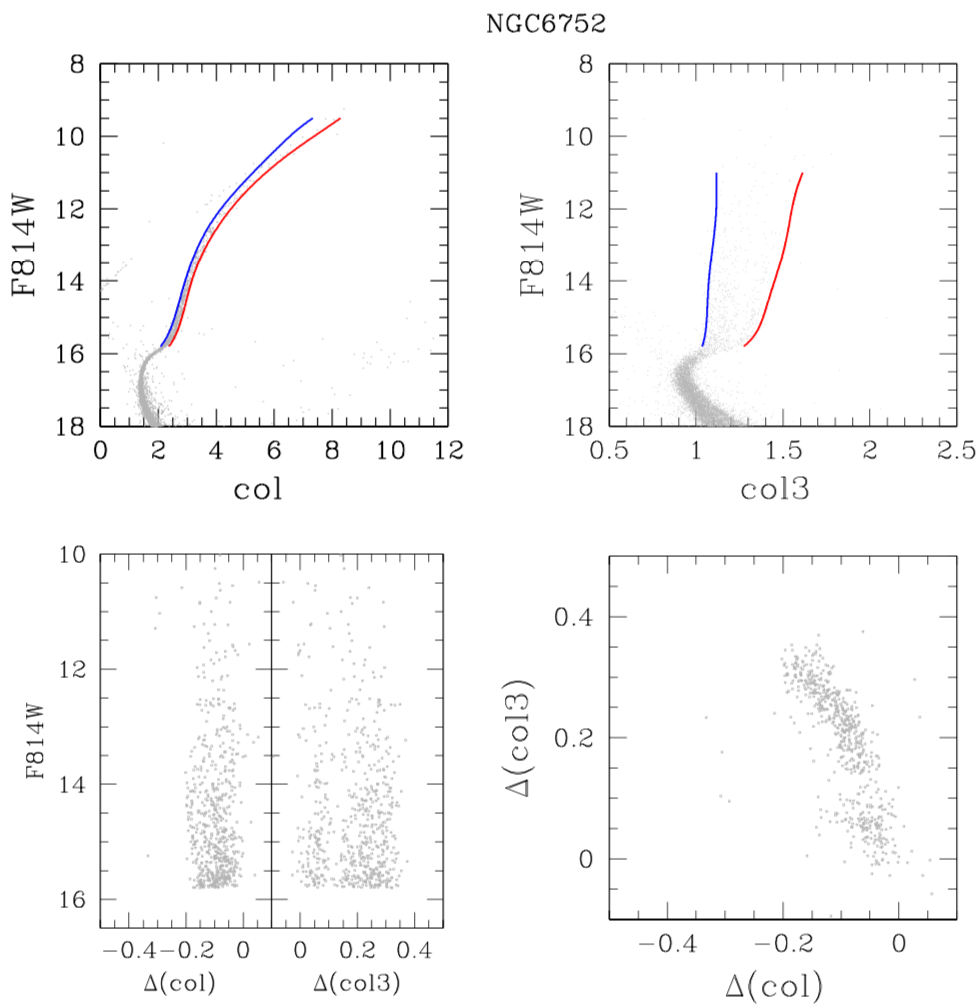}
\caption{As in Fig. 1 but for the globular cluster NGC~6752}
\label{f:app6752}
\end{figure*}

\begin{figure*}
\centering
\includegraphics[scale=1.1]{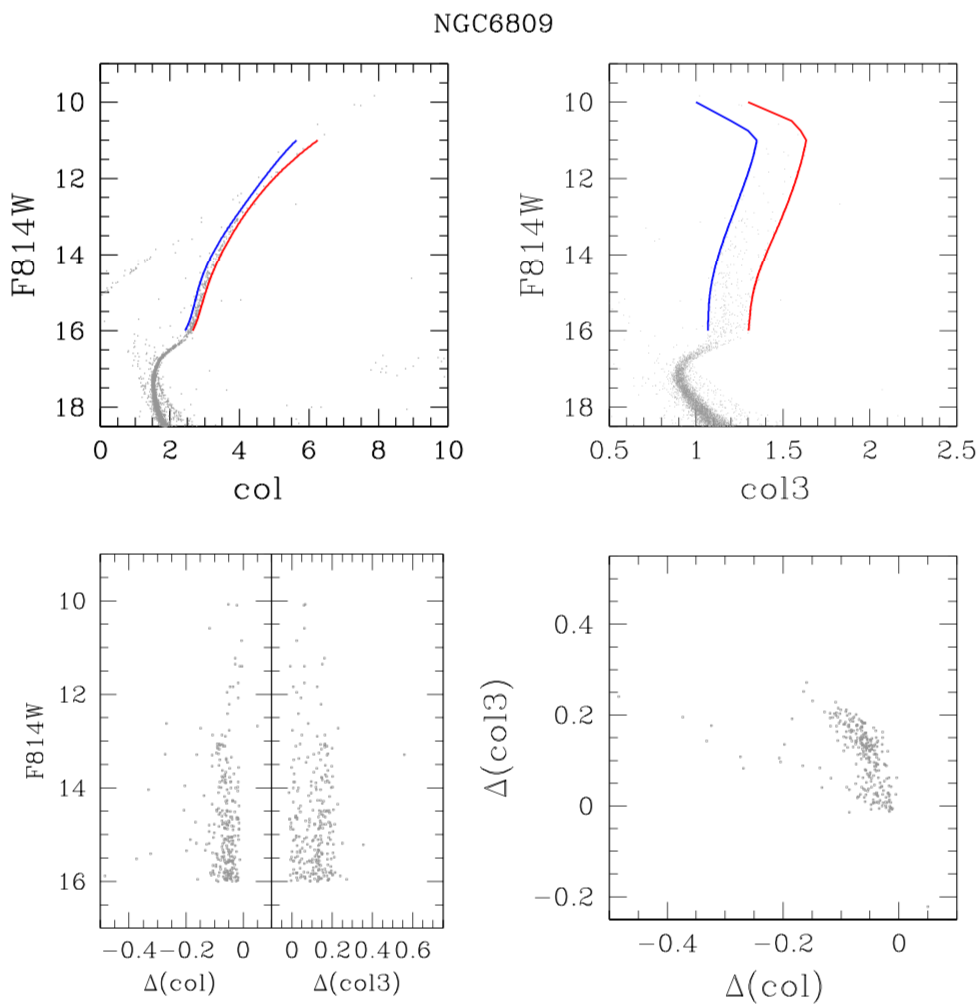}
\caption{As in Fig. 1 but for the globular cluster NGC~6809}
\label{f:app6809}
\end{figure*}

\begin{figure*}
\centering
\includegraphics[scale=1.1]{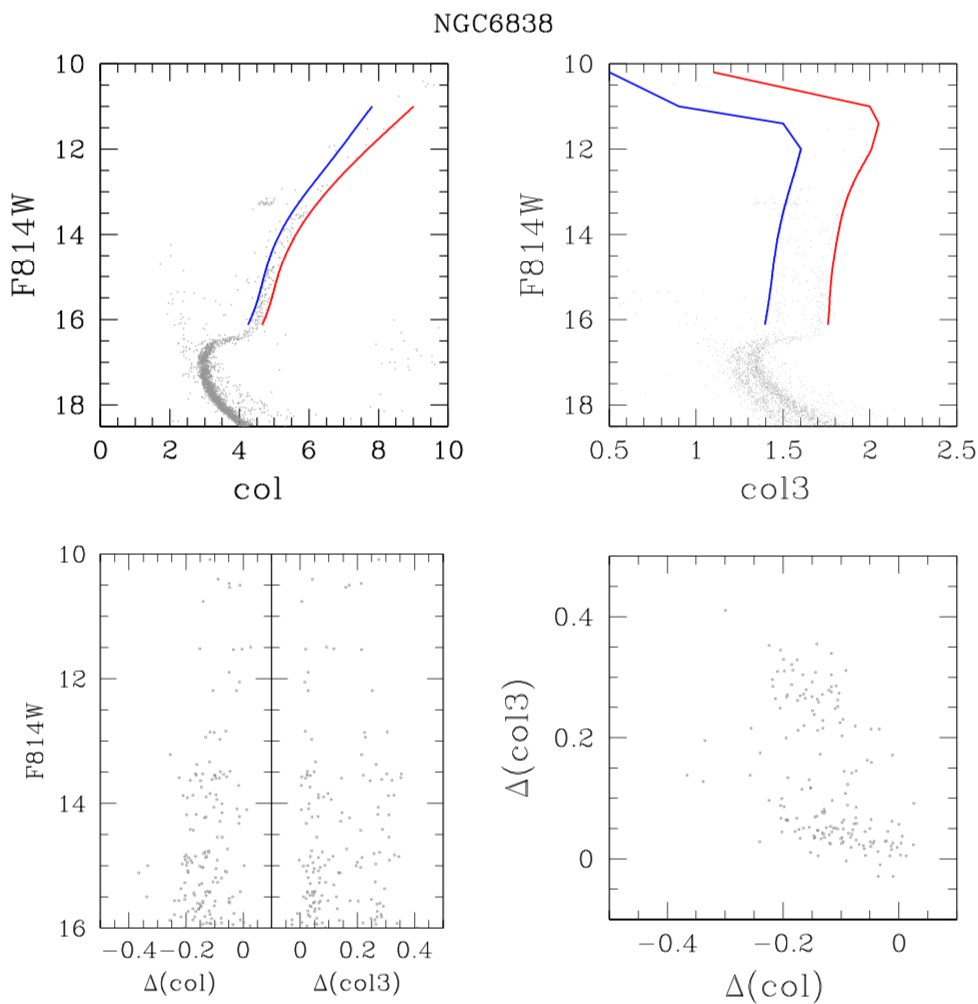}
\caption{As in Fig. 1 but for the globular cluster NGC~6838}
\label{f:6838}
\end{figure*}

\begin{figure*}
\centering
\includegraphics[scale=1.1]{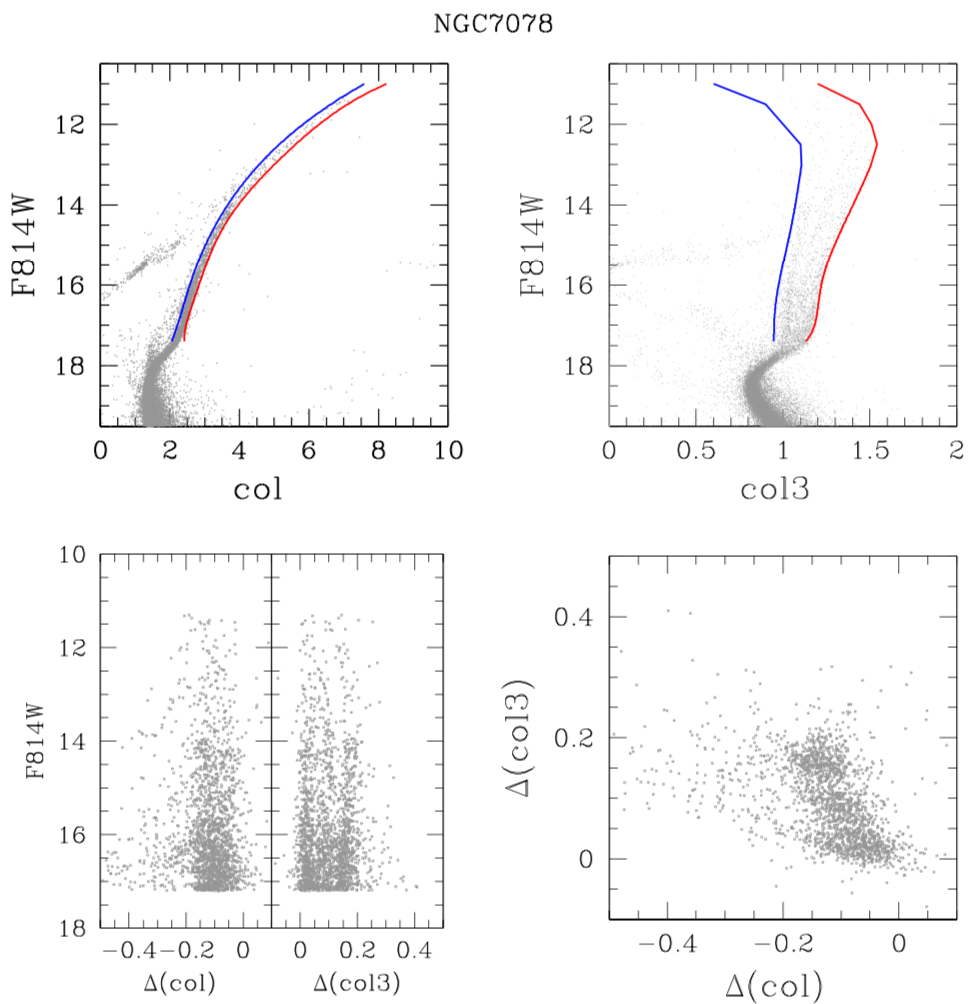}
\caption{As in Fig. 1 but for the globular cluster NGC~7078}
\label{f:app7078}
\end{figure*}

\begin{figure*}
\centering
\includegraphics[scale=1.1]{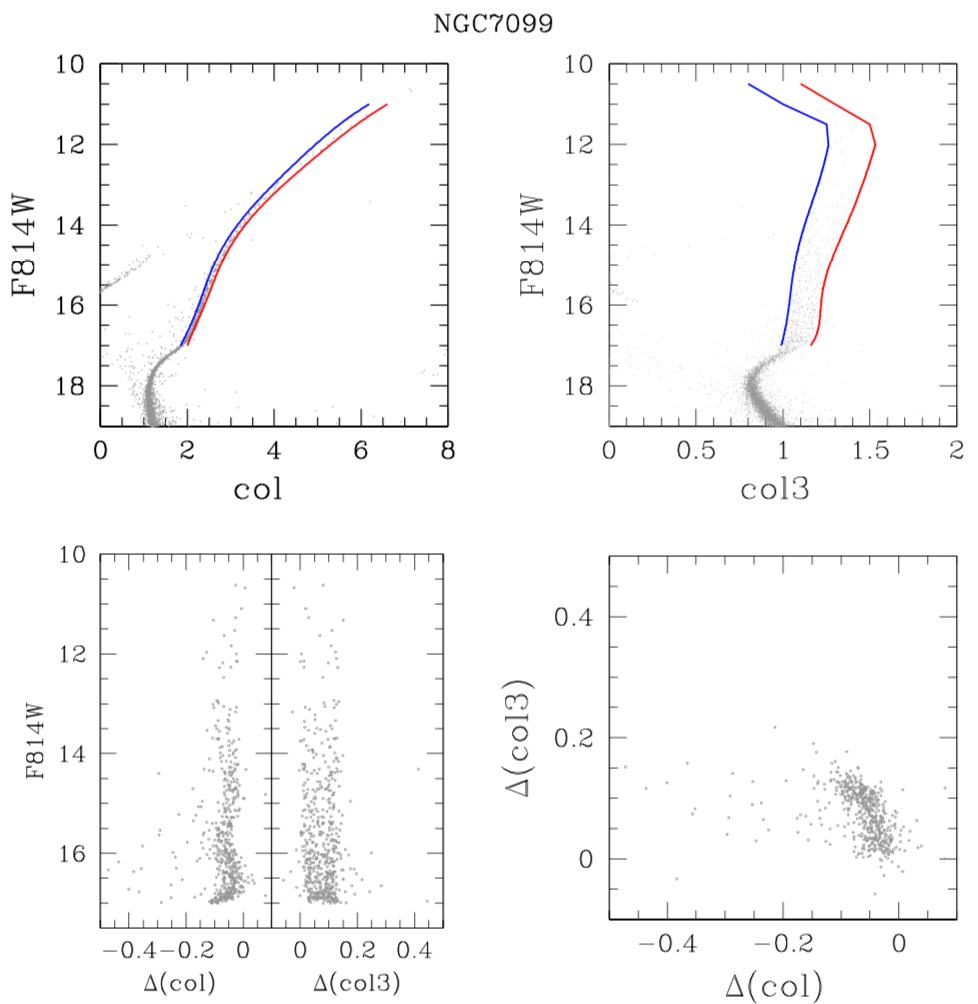}
\caption{As in Fig. 1 but for the globular cluster NGC~7099}
\label{f:app7099}
\end{figure*}

In Figures from Fig.~\ref{f:app0288} to Fig.~\ref{f:app7099} in the upper panels
are shown the CMDs $col$ versus $mag_{F814W}$
and $col3$ versus $mag_{F814W}$. The red and blue lines are the fiducials used
to normalise the pseudo-colours. In the lower left panels we show the
rectified RGBs in the normalized $col$ and $col3$. Finally, in the lower
right panel we present the resulting PCM.

Table A.1 lists all the PCMs derived in the present work. Star IDs are from the
catalogues of Nardiello et al. (2018) and are unique for each star within a
given GC. Here we only show an excerpt of this Table as a guidance. The whole
Table is available only in electronic form at CDS.

\begin{table*}[h]
\centering
\caption{Pseudo-colour maps (PCMs) for all 22 GCs in the sample. The complete
Table is only available at the CDS.}
\begin{tabular}{rrllrl}
\hline
$\Delta col$ & $\Delta col3$ & RA & Dec & star ID& GC \\
\hline
-0.2109 &   0.2258  &   6.076004  & -72.107040  & R0000586 & 104 \\
-0.2055 &   0.1613  &   6.066042  & -72.106667  & R0000641 & 104 \\
-0.1938 &   0.1755  &   6.067145  & -72.105949  & R0000766 & 104 \\
-0.1637 &   0.2712  &   6.058540  & -72.105705  & R0000811 & 104 \\
-0.1816 &   0.1747  &   6.047574  & -72.105133  & R0000910 & 104 \\
-0.1608 &   0.1571  &   6.023707  & -72.105064  & R0000917 & 104 \\
-0.0847 &  -0.0300  &   6.026685  & -72.105034  & R0000929 & 104 \\
-0.2605 &   0.2699  &   6.018813  & -72.104927  & R0000951 & 104 \\
-0.1281 &   0.2678  &   6.043403  & -72.104774  & R0000984 & 104 \\
-0.1612 &   0.0519  &   6.047808  & -72.104568  & R0001028 & 104 \\
-0.1510 &   0.1516  &   6.048434  & -72.104340  & R0001067 & 104 \\
-0.2553 &   0.2338  &   6.004899  & -72.104340  & R0001070 & 104 \\
 0.0521 &   0.4089  &   6.028763  & -72.104332  & R0001073 & 104 \\
-0.0756 &   0.0280  &   6.068101  & -72.104294  & R0001076 & 104 \\
-0.1837 &   0.2275  &   6.032630  & -72.104263  & R0001086 & 104 \\
-0.2091 &   0.1949  &   6.039308  & -72.104240  & R0001089 & 104 \\
-0.1940 &   0.2577  &   6.034832  & -72.104095  & R0001113 & 104 \\
-0.2905 &   0.2657  &   6.023297  & -72.104034  & R0001122 & 104 \\
-0.3772 &   0.3225  &   6.030381  & -72.104034  & R0001126 & 104 \\
-0.2396 &   0.2103  &   6.076389  & -72.103943  & R0001137 & 104 \\ 

\hline
\end{tabular}
\label{t:tabapp}
\end{table*}

\FloatBarrier

\newpage

\newpage

\newpage

\newpage

\newpage

\newpage

\newpage

\newpage

\newpage

\newpage

\newpage

\newpage

\newpage

\newpage

\newpage

\newpage

\newpage
\FloatBarrier

\newpage

\newpage

\newpage

\FloatBarrier
\newpage
\section{Notes on particular stars in a few GCs}

In a few plots some stars apparently are either missing with respect to the
numbers given in Table~\ref{t:numbers} or are  in an odd position on the PCMs.
In the following we highlight possible problems related to the identification of
stars with spectroscopic abundances from the figures in Marino et al. (2019) and
we note about particular stars on the PCMs. Flag F indicates stars found
and used by Marino et al. (2019) and flag M indicates additional stars found
only by us (see Section 2.2).

\paragraph{NGC~104} Star R0009274 could be a star on the AGB, according to its
position in the CMD mag$_{F275W}$, (mag$_{F275W}$-mag$_{F814W})$ (although not
judging from other CMDs).

\paragraph{NGC~362} Stars R0000796 and R0009389 (both found also in Marino et
al. 2019) could be among the so called red-RGB stars, according to the 
mag$_{F336W}$, (mag$_{F336W}$-mag$_{F814W})$ CMD. To these, we add star
R0009282, neglected by Marino et al., as a possible red-RGB star.

\paragraph{NGC~1851} Two stars neglected in Marino et al. (2019, 
star R0000480 and star R0015479) are located in the region attributed to red-RGB
stars, the same of star R0015322, used by Marino et al.

\paragraph{NGC~2808} There is a star in the PCM by Marino et al. (2019) located
at about $\Delta col, \Delta col3 = (-0.416, 0.175)$ and classified as FG star
(figure 1 in Marino et al.). We note that this is the object making  the
sequence of FG stars in NGC~2808 so extended. However, we were unable to find  a star located at this value of $\Delta col3$ in
their figure 9, even if all the
matches in this GC have a Na abundance. For unknown reasons, this star, plotted
on their PCM as having spectroscopic abundances, was then dropped in the
Na-$\Delta col3$ relation. Several stars in this GC, both found and neglected in
Marino et al. (e.g. stars R0033484, R0027893, R0016679, R0009902) seem to lie in
the saturation region. Stars R0005798 (flag=F) and star R0021965 (flag=M) are
outside the RGB in the mag$_{F336W}$, (mag$_{F336W}$-mag$_{F814W})$,  but not in
the mag$_{F275W}$, (mag$_{F275W}$-mag$_{F814W})$.

\paragraph{NGC~6121} On the Na-$\Delta col3$ relation derived by us, three stars
with flag=M (R0000373, R0000417, and R0000781; [Na/Fe]=0.05, 0.41, and 0.43) are
superimposed to stars R0000301, R0000863, and R0000749 ([Na/Fe]=0.06, 0.38, and
0.40), flag=F,  respectively. Star R0001022 ([Na/Fe]=0.25) is superimposed to
star R0001728 ([Na/Fe]=0.24), both with flag=F. Judging by its position on the
PCM, star R0000101 (flag=F) could be a mis-identification of ours. Both the
superpositions and the possible mis-identification are mostly due to the
abundances in Marino et al. (2008) being provided with only 2 significant
decimal figures, so that many stars result with the same or very similar Na and are difficult to assign only on the base of the figures in Marino et al. (2019).

\paragraph{NGC~6205} In this GC we identified  three stars from the figures in Marino et al.
(2019) that do not have the complete  set
of colours to make them appear on the PCM (based on the photometry by Nardiello et al. 2018), even if they are likely matches according to the coordinates.
Star R0001523 (L-666, [Na/Fe]=0.12 in Johnson and Pilachowski 2012) does
not have mag$_{F336W}$. Star R0012534 (L-261, [Na/Fe]=0.06 in Johnson and
Pilachowski) does not have mag$_{F336W}$. Star R0002021 ( L-324,
[Na/Fe]=0.27 in Johnson and Pilachowski) does not have mag$_{F438W}$. It is
unlikely that we mis-identified all three: e.g. among the stars matched in
coordinates between Nardiello et al. and Johnson and Pilachowski there is no
star having [Na/Fe]=0.12, such as star L-666=R0001523. When considering both
coordinates and Na abundances we have no explanation on why these objects are
located on the PCM used in Marino et al. (2019). Finally, star R0001319 (L-370,
[Na/Fe]=0.27), with flag=F is superimposed to star R0003036 (L=296,
[Na/Fe]=0.28), with flag=M.

\paragraph{NGC~6254} Star R0001704 ([Na/Fe]=-0.137, flag=F) is superimposed to
star R0001056 ([Na/Fe]=-0.128, flag=M).

\paragraph{NGC~6715} Star R257062 ([Na/Fe]=0.587) found by Marino et al. (2019)
should not be on a PCM, having no mag$_{F814W}$. Perhaps this is a mis-identification 
in Marino et al., confused for another star. 

\paragraph{NGC~6752} Analogously, star R010792, found in Marino et al. (2019)
should not be in a PCM, having no mag$_{F336W}$. From the Na-O anti-correlation
shown in Figure 8 of Marino et al. this object is unambiguously identified as star
NGC~6752-21694 in Carretta et al. (2009a), with ([O/Fe], [Na/Fe]=0.091, 0.318),
yet we were not able to put it on our PCM for the lack of all required 
magnitudes. It is unclear how this star can appear among those with 
both HST photometry and Na abundances in Marino et al. (2019).

\end{appendix}


\begin{thebibliography}{}

\bibitem[]{} Bastian, N., Lardo, C. 2018, ARA\&A, 56, 83
\bibitem[]{} Bragaglia, A., Carretta, E., D'Orazi, V. et al. 2017, A\&A, 607, A44 
\bibitem[]{} Cadelano, M., Dalessandro, E., Vesperini, E. 2024, arXiv:2402.09514
\bibitem[]{} Calamida, A., Bono, G., Stetson, P.B. et al. 2007, ApJ, 670, 400
\bibitem[]{} Carretta, E. 2006, AJ, 131, 1766 
\bibitem[]{} Carretta, E. 2013, A\&A, 557, A128 
\bibitem[]{} Carretta, E. 2015, ApJ, 810, 148 
\bibitem[]{} Carretta, E. 2016, IAUS 317, 97
\bibitem[]{} Carretta, E., Bragaglia, A. 2023, A\&A, 677, A73 
\bibitem[]{} Carretta, E., Bragaglia, A., Gratton, R.G., Lucatello,
  S. Momany, Y. 2007, A\&A, 464, 927 
\bibitem[]{}  Carretta, E., Bragaglia, A., Gratton, R.G. et al. 2009a, 
  A\&A, 505, 117 
\bibitem[]{} Carretta, E., Bragaglia, A., Gratton, R.G., Lucatello, S. 2009b, 
 A\&A, 505, 139 
\bibitem[]{} Carretta, E., Bragaglia, A., Gratton, R.G., D'Orazi, V., Lucatello,
 S. 2009c, A\&A, 508, 695 
\bibitem[]{} Carretta, E., Bragaglia, A., Gratton, R.G. et al. 2010a, A\&A, 516, 55 
\bibitem[]{} Carretta, E., Bragaglia, A., Gratton, R.G. et al. 2010b, A\&A, 520, 95 
\bibitem[]{} Carretta, E., Bragaglia, A., Gratton, R.G., D'Orazi, V., 
  Lucatello, S. 2011a, A\&A, 535, 121 
\bibitem[]{} Carretta, E., Lucatello, S., Gratton, R.G., Bragaglia, A., D'Orazi,
  V. 2011b, A\&A, 533, 69 
\bibitem[]{} Carretta, E., Bragaglia, A., Gratton, R.G. et al. 2013,
  A\&A, 557, A138 
\bibitem[]{} Carretta, E., Bragaglia, A., Gratton, R.G. et al. 2014, A\&A, 564, A60 
\bibitem[]{} Carretta, E., Bragaglia, A., Gratton, R.G. et al. 2015, A\&A, 578, A116 
\bibitem[]{} Charbonnel, C. 1994, A\&A, 282, 811
\bibitem[]{} Cohen, J.G., Briley, M.M., Stetson, P.B. 2002, AJ, 123, 2525 
\bibitem[]{} Cordero, M.J., Pilachowski, C.A., Johnson C.I. et al. 2014, ApJ, 780, 94 
\bibitem[]{} Cottrell, P.L., Da Costa, G.S. 1981, ApJL, 245, L79
\bibitem[]{} Dalessandro, E., Cadelano, M., Vesperini, E. et al. 2019, ApJ, L24
\bibitem[]{} Decressin, T., Meynet, G., Charbonnel C. Prantzos, N.,
 Ekstrom, S. 2007, A\&A, 464, 1029 
\bibitem[]{} de Mink, S.E., Pols, O.R., Langer, N., Izzard, R.G. 2009, A\&A,
  507, L1 
\bibitem[]{} Denissenkov, P.A., Hartwick, F.D.A. 2014, MNRAS, 437, L21 
\bibitem[]{} D'Ercole, A., D'Antona, F., Ventura, P., Vesperini, E., McMillan,
  S.L.W. 2010, MNRAS, 407, 854 
\bibitem[]{} Gratton, R.G., Sneden, C., Carretta, E., Bragaglia, A. 2000, 
  A\&A, 354, 169 
\bibitem[]{} Gratton, R.G., Bonifacio, P., Bragaglia, A., et al. 2001, 
 A\&A, 369, 87 
\bibitem[]{} Gratton, R.G., Sneden, C., Carretta, E. 2004, ARA\&A, 42, 385
\bibitem[]{} Gratton, R.G., Bragaglia, A., Carretta, E., D'Orazi, V., 
  Lucatello, S., Sollima, A. 2019, A\&ARv, 27, 8
\bibitem[]{} Johnson, C.I., Pilachowski, C.A. 2012, ApJ, 754, L38 
\bibitem[]{} Johnson, C.I., Calamida, A., Kader, J.A. et al. 2023, AJ, 163, 3
\bibitem[]{} Koch, A., Grebel, E.K., Martell, S.L. 2019, A\&A, 625, A75 
\bibitem[]{} Kraft, R.~P. 1994, PASP, 106, 553 
\bibitem[]{} Langer, G.E., Hoffman, R., Sneden, C. 1993, PASP, 105, 301
\bibitem[]{} Larsen, S.S., Brodie, J.P., Grundahl, F., Strader, J. 2014, ApJ,
  797, 15 
\bibitem[]{} Lee, J.-W. 2019, ApJ, 883, 166
\bibitem[]{} Lee, J.-W. 2023, ApJ, 961, 227
\bibitem[]{} Lee, J.-W., Lee, J., Kang, Y.-W. et al. 2009, ApJ, 695, L78 
\bibitem[]{} Marino, A.F., Villanova, S., Piotto, G. et al. 2008, A\&A, 490, 625 
\bibitem[]{} Marino, A.F., Milone, A.P., Renzini, A. et al. 2019, MNRAS, 487,
  3815 
\bibitem[]{} Martell, S.L., Smolinski, J.P., Beers, T.C., Grebel, E.K. 2011,
  A\&A, 534, 136 
\bibitem[]{} Massari, D., Lapenna, E., Bragaglia, A. et al. 2016, MNRAS, 458,
  4162 
\bibitem[]{} M\'esz\'aros, S., Masseron, T., Garc\'ia-Hern\'andez, D.A. et al.
  2020, MNRAS, 492, 1641 
\bibitem[]{} Milone, A.P., Piotto, G., Renzini, A. et al. 2017, MNRAS, 464,
  3636 
\bibitem[]{} Monelli, M., Milone, A.P., Stetson, P.B. et al. 2013, MNRAS, 431,
  2126 
\bibitem[]{} Nardiello, D., Libralato, M., Piotto, G. et al. 2018, MNRAS, 481,
  3382 
\bibitem[]{} Pancino, E., Romano, D., Tang, B. et al. 2017. A\&A, 601, A112  
\bibitem[]{} Piotto, G., Milone, A.P., Bedin, L.R. et al. 2015, AJ, 149, 91   
\bibitem[]{} Prantzos, N., Charbonnel, C. 2006, A\&A, 458, 135 
\bibitem[]{} Prantzos, N., Charbonnel, C., Iliadis, C. 2017, A\&A, 608, 
 A28 
\bibitem[]{} Sbordone, L., Salaris, M., Weiss, A., Cassisi, S. 2011, A\&A, 534,
  A9 
\bibitem[]{} Smith, G.H. 1987, PASP, 99, 67 
\bibitem[]{} Smith, G.H. 2015, PASP, 127, 825 
\bibitem[]{} Smith, G.H., Briley, M.M. 2005, PASP, 117, 895 
\bibitem[]{} Smith, G.H., Briley, M.M. 2006, PASP, 118, 740 
\bibitem[]{} Smith, G.H., Modi, P.N., Hamren, K. 2013, PASP, 125, 1287
\bibitem[]{} Taylor, M.B. 2005, Astronomical Data Analysis Software and Systems
  XIV, 347, 29
\bibitem[]{} Vargas, C., Villanova, S., Geisler, D. et al. 2022, MNRAS, 515,
  1903
\bibitem[]{} Ventura, P. D'Antona, F., Mazzitelli, I., Gratton, R. 2001,
  ApJ, 550, L65 
\bibitem[]{} Vesperini, E., McMillan, S.L.W., D'Antona, F., D'Ercole, A. 2010,
  ApJ, 718, 112 
\bibitem[]{} Vesperini, E., McMillan, S.L.W., D'Antona, F., D'Ercole, A. 2013,
  MNRAS, 429, 1913 

\end{thebibliography}
\end{document}